\documentclass[aps,prd,twocolumn,nofootinbib,showpacs,floatfix]{revtex4}
\usepackage{amsmath, amsthm}
\usepackage[dvips]{graphicx}

\newcommand{\lsim}{\lower0.6ex\vbox{\hbox{$ \buildrel{\textstyle <}\over{\sim}\ $}}}
\newcommand{\gsim}{\lower0.6ex\vbox{\hbox{$ \buildrel{\textstyle >}\over{\sim}\ $}}}
\newcommand{\beq}{\begin{equation}}
\newcommand{\eeq}{\end{equation}}

\newcommand{\Pktom}[2]{P_{\kappa}^{\mathrm{#1}\mathrm{#2}}}

\newcommand{\wlw}[1]{W_{\mathrm{#1}}}
\newcommand{\Da}{D_{\mathrm{A}}}

\newcommand{\ntomo}{N_{\mathrm{TOM}}}

\newcommand{\dd}{\mathrm{d}}

\begin{document}


\title{
Weak Gravitational Lensing as a Method to Constrain Unstable Dark Matter
}

\author{Mei-Yu Wang and Andrew R. Zentner}
\affiliation{
Department of Physics and Astronomy, University of Pittsburgh, 
Pittsburgh, PA 15260
}
\email{
mew56+@pitt.edu,zentner@pitt.edu
}


\date{\today}


\begin{abstract}

The nature of the dark matter remains a mystery. The possibility 
of an unstable dark matter particle decaying to invisible daughter 
particles has been explored many times in the past few decades.  
Meanwhile, weak gravitational lensing shear has gained a lot 
of attention as probe of dark energy, though it was previously 
considered a dark matter probe.  
Weak lensing is a useful tool for constraining the 
stability of the dark matter. In the coming decade a number 
of large galaxy imaging surveys will be undertaken and will 
measure the statistics of cosmological weak lensing with unprecedented 
precision. Weak lensing statistics are sensitive to unstable dark matter 
in at least two ways.  Dark matter decays alter the matter 
power spectrum and change the angular diameter distance-redshift 
relation. We show how measurements of weak lensing shear correlations 
may provide the most restrictive, model-independent constraints on the 
lifetime of unstable dark matter.  Our results rely on assumptions 
regarding nonlinear evolution of density fluctuations in scenarios 
of unstable dark matter and one of our aims is to stimulate interest 
in theoretical work on nonlinear structure growth in unstable dark 
matter models.

\end{abstract}



\maketitle

\section{Introduction}
\label{section:introduction}

A preponderance of evidence supports a picture in which $\sim 5/6$ of the mass density of the Universe 
resides nonbaryonic dark matter 
(reviews include Refs.~\citep{jungman_etal96b,griest_kamionkowski00,bertone_etal05}).  
The prevailing hypothesis is that the dark matter is an as yet undetected particle that survives 
as a relic from the hot, early Universe.  An effort to identify the dark matter now proceeds on 
many fronts and the dark matter is currently constrained by direct searches 
(e.g., Refs.~\citep{cdms10,xenon10,dama10,cogent10,xenon100}), 
indirect searches (e.g., Refs.~\citep{icecube09,fermi10}), and astronomical observations 
(e.g., Refs.~\citep{abazajian_etal07,scott_etal08,boyarsky_etal08,Komatsu_etal09}).  
In this paper, we explore the possibility of constraining invisible decays 
of the dark matter particle using forthcoming statistical measurements
of weak gravitational lensing.

Limits on unstable dark matter have been considered by numerous authors in the recent literature 
\citep{Kaplinghat_etal99,zentner_walker02,ichiki_etal04,Oguri_etal03,peutzfeld_chen04,palomares_ruiz08,
borzumati_etal08,Gong_etal08,Amigo_etal08,peter10,peter_etal10}.  Radiative decays are very 
strictly limited, with the best constraints yielding lifetime bounds \
of $\tau_{\rm DDM} \gsim 10^7 H_0^{-1}$ 
\citep{chen_kamionkowski04,hansen_haiman04,mapelli_etal06,zhang_etal07,yuksel_kistler08}.  
Assuming decays to Standard Model neutrinos, the least detectable Standard Model particles, places 
mass-dependent limits as restrictive as $\tau_{\rm DDM} \gsim 10^{3} H_0^{-1}$, for dark matter 
particle masses near $10^2$~GeV \citep{palomares_ruiz08,feldstein_fitzpatrick10}, though 
constraints are strongly mass dependent.  
Cosmological tests provide an opportunity to constrain the stability of the 
dark matter independent of particle mass and the interactions of the decay products, 
and current cosmological limits on invisible dark matter decays imply that the 
lifetime for decays to relatively light products is $\tau_{\rm DDM} \gsim 50 H_0^{-1}$ 
\citep{zentner_walker02,Oguri_etal03,Takahashi_etal04,Gong_etal08,Amigo_etal08,borzumati_etal08}.

We consider the possibility of improving model-independent constraints 
on the dark matter particle lifetime using forthcoming weak lensing data.  
Such independent constraints would be most relevant to models of light 
dark matter (masses $\lsim 10$~MeV to evade neutrino constraints) 
or dark matter sufficiently sequestered from the Standard Model 
(e.g., Refs.~\citep{feng_etal08,feng_etal09,ackerman_etal09,peter10,peter_etal10}) 
and may help improve or complement constraints of asymmetric dark matter 
models \citep{falkowski_etal09,kaplan_etal09,feldstein_fitzpatrick10}.  
For concreteness, we consider constraints on dark matter lifetime 
in a benchmark model of a cold dark matter particle that undergoes 
two-body decay to light daughter particles with a 
lifetime tuned to be exceptionally large.

The primary constraint comes from scale- and redshift-dependence 
of the cosmological gravitational lensing power spectrum at $z \sim 0-3$ after 
normalizing the power spectrum of density fluctuations at $z \approx 1100$ via 
cosmic microwave background (CMB) measurements.  
A full exploration of possible constraints 
is difficult because the nonlinear evolution of 
structure in the Universe in such models has not been 
extensively studied.  In our most conservative forecasts for 
what may be possible with forthcoming instruments, we find that utilizing 
only scales on which linear perturbative evolution should be useful ($\ell < 300$) 
and taking weak prior constraints on other cosmological parameters, 
forthcoming surveys may produce dark matter lifetime constraints that 
are, at minimum, competitive with contemporary, model-independent constraints.  
More aggressive priors expected from Planck CMB measurements improve upon 
these constraints by roughly a factor of two. 
We argue that utilizing the information that may be available 
from nonlinear evolution should improve upon these constraints 
by an order of magnitude.  Our most aggressive forecasts, 
using information extending to scales $\ell \le 3000$ and 
taking Planck CMB prior constraints on other cosmological 
parameters, suggest that constraints as strong as 
$\tau_{\rm DDM} \gsim \mathrm{a}\ \mathrm{few}\ \times 10^{2} H_0^{-1}$ may be possible 
with a survey covering a large fraction of the sky, 
such as that planned by the Large Synoptic Survey Telescope (LSST) 
\citep{lsstbook} or Euclid \citep{eicbook}.  Achieving reliable constraints from such 
measurements requires an understanding of nonlinear evolution in unstable 
dark matter models.  The possibility of stringent constraints 
on decaying dark matter from astronomical imaging surveys should be 
strong motivation to study the nonlinear evolution of cosmic 
structure formation in such scenarios (see also Refs.~\citep{peter10,peter_etal10}).

We continue our manuscript with a discussion of weak lensing 
observables in \S~\ref{section:Weak Lensing}.  We describe 
our benchmark model for unstable dark matter and the 
evolution of cosmological perturbations in such a 
model in \S~\ref{section:model}.  We discuss nonlinear 
structure evolution and the two methods we use to estimate 
nonlinear evolution in \S~\ref{section:Nonlinear}.  
We describe our methods for forecasting constraints on unstable dark 
matter in \S~\ref{section:fisher}.  
This section also includes a summary of our fiducial 
cosmological model and our assumptions regarding prior 
constraints.  We illustrate the effects of 
unstable dark matter on lensing observables and 
present our forecast limits on dark matter lifetimes 
in \S~\ref{section:results}.  Finally, we summarize and 
discuss avenues for future work in \S~\ref{section:conclusions}.

\section{Weak Gravitational Lensing Observables}
\label{section:Weak Lensing}

We explore the utility of weak gravitational lensing measurements for 
constraining the stability of the dark matter.  Our most robust 
forecasts derive from considerations of possible weak lensing 
measurements restricted to scales where linear perturbative 
evolution of the metric potentials remains useful.  In this manner, 
our paper is very similar in spirit to that of \citet{schmidt08}, 
who studied constraints on modified gravity from weak lensing 
statistics restricted to linear scales.  However, we attempt to 
estimate possible improvements to the constraining power 
of weak lensing observables, provided that nonlinear evolution can be 
modeled robustly.

We consider the set of observables that may be available from ongoing and 
forthcoming large-scale galaxy imaging surveys to be the auto- and 
cross-spectra of lensing convergence from sets of galaxies in 
$\ntomo$ redshift bins.  The $\ntomo (\ntomo+1)/2$ distinct 
convergence spectra are
\beq
\label{eq:pkij}
\Pktom{i}{j}(\ell) =\ell^4 \int \dd z  \frac{\wlw{i}(z)\wlw{j}(z)}{H(z) D_A^6(z)}P_{\Psi-\Phi}(k=\ell/\Da,z),
\eeq
where $i$ and $j$ label the redshift bins of the source galaxies.  We take 
$\ntomo=5$ and consider evenly-spaced bins in redshift from a minimum redshift 
of $z=0$ to a maximum redshift of $z=3$.  In agreement with the study of 
\citet{ma_etal06}, we find that finer binning is not required to maximize 
the constraining power of such surveys.  Weak lensing as a cosmological probe 
has been discussed at length in numerous papers 
(a recent review is Ref.~\citep{huterer10}).  We give a brief 
description of our methods below, which are based on the 
conventions and notation in Ref.~\citep{zentner_etal08} 
(to which we refer the interested reader for details).

In Eq.~\ref{eq:pkij}, 
$H(z)$ is the Hubble expansion rate, 
$D_A$ is the comoving angular diameter distance, 
and  $P_{\Psi-\Phi}(k,z)$ is the power spectrum of 
Newtonian gauge scalar potentials $\Psi-\Phi$ at 
wavenumber $k$ and redshift $z$.  
In the following section, we describe our use of the 
publicly-available {\tt CMBFAST} code to calculate
$P_{\Psi-\Phi}(k,z)$, in which case it will be more natural 
to work in the synchronous gauge.  Transforming between 
coordinate systems can be accomplished straightforwardly by 
following, for example, the methods described 
in Ref.~\citep{ma_bertschinger95} which we do not repeat here.

The $W_{i}$ are the so-called lensing weight functions for source galaxies 
in redshift bin $i$.  In practice, the galaxies will be binned by 
photometric redshift, so that the bins will have nontrivial overlap 
in true redshift (see Ref.~\citep{ma_etal06} for a detailed discussion).  
Defining the true redshift distribution of source galaxies in the 
$i$th photometric redshift bin as $dn_i/dz$, the window functions are 
\beq
\label{eq:W}
W_{i}(z)=D_A \int dz' {D_A(z,z') \over D_A(z')} {dn_i \over dz'}
\eeq
where $D_A(z,z')$ is the angular diameter 
distance between redshift $z$ and $z'$.

We model the increased uncertainty induced by utilizing photometric galaxy redshifts 
with the probability function of assigning an individual source galaxy photometric 
redshift $z_p$ given a true redshift $z$, $P(z_p|z)$.  In this notation, the true 
redshift distribution of sources in the $i$th photometric redshift bin is
\beq
\label{eq:dnidz}
\frac{dn_i(z)}{dz} = \int^{z_{p,i}^{(high)}}_{z_{p,i}^{(low)}}\ dz_p\ \frac{dn(z)}{dz}\ P(z_p|z)
\eeq
Here we take the true redshift distribution to be 
\beq
\label{eq:dndz}
{dn(z) \over dz}=\bar{n} {4z^2 \over \sqrt{2 \pi z_0^3}} \exp[-(z/z_0)^2]
\eeq
with $z_0 \simeq 0.92$, so that the median survey redshift to $z_{med}$ = 1, 
and $\bar{n}$ as the total density of source galaxies per
unit solid angle \citep{smail_etal95a,smail_etal95b,newman08}.  
We assume that uncertain photometric redshifts can be approximated 
by taking
\beq
\label{eq:p}
P(z_p|z)={1 \over \sqrt{2 \pi \sigma_z}} \exp\left[-{(z_{p}-z)^2 \over 2\sigma_z^2}\right]
\eeq
where $\sigma_z(z)=0.05(1+z)$ \citep{ma_etal06}. Complexity 
in photometric redshift distributions is something that must 
be overcome to bring weak lensing constraints on cosmology 
to fruition (e.g., Ref.~\citep{bernstein_huterer10,hearin_etal10}).

Observed convergence power spectra $\bar{P}^{ij}_{\kappa}(\ell)$, 
contain both signal and shot noise,
\beq
\label{trueP}
\bar{P}^{ij}_{\kappa}(\ell)=P_{\kappa}^{ij}+n_{i}\delta_{ij}\langle \gamma^2 \rangle 
\eeq
where $\langle \gamma^2 \rangle$ is the noise from 
intrinsic ellipticities of source galaxies, 
and $n_i$ is the surface density of galaxies in the 
ith tomographic bin.  We follow the recent convention 
and set $\sqrt{\langle \gamma^2 \rangle}=0.2$, subsuming 
additional errors on galaxy shape measurements into an 
effective mean number density of galaxies, $\bar{n}$.  
Assessments of intrinsic shape noise per galaxy may 
be found in, for example \cite{massey_etal04,kasliwal_etal08,lsstbook}.  
Assuming Gaussianity of the lensing field, the covariance between 
observables $\bar{P}^{ij}_{\kappa}$ and $\bar{P}^{kl}_{\kappa}$ is
\beq
\label{cab}
C_{AB}=\bar{P}^{ik}_{\kappa}\bar{P}^{jl}_{\kappa}+\bar{P}^{il}_{\kappa}\bar{P}^{jk}_{\kappa}
\eeq
where the i and j map to the observable index A, and k and l map to B such that 
$C_{AB}$ is a square covariance matrix with $\ntomo (\ntomo+1)/2$ rows and columns.  
We assume Gaussianity throughout this work and consider only multipoles 
$\ell < 3000$ at which point the Gaussian assumption and several weak lensing 
approximations break down 
\cite{white_hu00,cooray_hu01,vale_white03,dodelson_etal06,semboloni_etal06}.

\section{Perturbation Theory with Unstable Dark Matter}
\label{section:model}

Our aim is to predict the power spectrum of weak gravitational 
lensing convergence in unstable dark matter scenarios.  
To do so, we must compute the modifications to the metric 
potentials in unstable dark matter scenarios [see Eq.~(\ref{eq:pkij})].  
We explore a restricted set of models in which a 
massive parent dark matter particle decays into 
a significantly lighter pair of daughter particles.  
For the sake of specificity, we adopt a 
decaying dark matter (DDM) scenario in which 
massive majorana parent particles decay 
into relativistic daughter (RD) particles via 
two-body decay and use this scenario to 
benchmark observational constraints.  
In such a scenario, the lifetime of the unstable dark matter 
particle lifetime ($\tau$) is the only nonstandard free parameter.  
One could assume decay to a combination of heavy and light 
daughter particles in which case the mass differences 
are important additional parameters that establish the 
recoil velocities of the decay product particles 
(as explored in detail in Refs.~\cite{peter10,peter_etal10}, recently).

The distribution functions of DDM ($f_{\rm DDM}$) and 
RD ($f_{\rm RD}$) evolve according to 
the coupled Einstein-Boltzmann equations.  In particular 
(e.g., \citep{ma_bertschinger95,Kaplinghat_etal99}), 
\begin{align}
{df_{\rm DDM} \over d \tau}& =-{a^2m_{\rm DDM} \over \tau_{\rm DDM}\epsilon_{\rm DDM}}f_{\rm DDM} \simeq -{1\over \tau_{\rm DDM}}f_{\rm DDM} \label{f_ddm}\\
{df_{\rm RD} \over d\tau} & =\;{a^2m_{\rm DDM} \over \tau_{\rm DDM}\epsilon_{\rm DDM}}f_{\rm DDM} \simeq {1\over \tau_{\rm DDM}}f_{\rm DDM} \label{f_rd},
\end{align}
where $\tau$  is the conformal time and $\tau_{DDM}$, $\epsilon_{DDM}$, and $m_{DDM}$ 
are the lifetime, energy, and mass of decaying dark matter.
Following established procedure, 
we express the distribution function of species $\rm X$ 
as a zeroth-order distribution plus a perturbation, 
\beq
\label{eq:perturb}
f_{\rm X}(\vec{x},\vec{q},\tau)=f_{\rm X}^0(q,\tau)[1+\Psi_{\rm X}(\vec{x},\vec{q},\tau)]
\eeq
The evolution of the mean energy density for DDM and its RD particles follow from 
the zeroth-order integrals of Eq.~(\ref{f_ddm}) and Eq.~(\ref{f_rd}), 
\begin{align}
\dot{\rho}_{\rm DDM}+3H\rho_{\rm DDM}&=-\Gamma\rho_{\rm DDM} \label{rho_ddm}\\
\dot{\rho}_{\rm RD}+4H\rho_{\rm RD}&=\Gamma\rho_{\rm DDM} \label{rho_rd}
\end{align}
Here and throughout, we designate $\dot{y}$ as the time derivative of $y$, and we 
denote the decay rate as $\Gamma=1/\tau_{\rm DDM}$.  In the limit of a massive 
DDM particle, evolution of the comoving density $\rho_{\rm DDM} a^3$ approaches 
$\propto \exp(-t/\tau_{\rm DDM})$.

The collision term describing the DDM decays is proportional to 
$f^0_{\rm DDM}$, rendering the equations describing the evolution of 
DDM perturbations identical to those of standard, stable cold dark matter at 
the lowest order in perturbation theory.  The perturbation equations describing 
the daughter particles are less trivial.  Following Refs.~\citep{Kaplinghat_etal99,ma_bertschinger95}, 
we expand the perturbation equations for RD particles in a series of 
Legendre polynomials $P_l(x)$, yielding 
\begin{multline}
\label{eq:F_RD1}
F_{\rm RD}(\vec{k},\hat{n},\tau) \;=\; {\int dq q^3 f^0_{\rm RD}(q,\tau)\Psi_{\rm RD} \over \int dq q^3  f^0_{\rm RD}(q,\tau)} \\
=\sum^{\infty}_{l=0}  (-i)^l(2l+1)F_{{\rm RD},l}(\vec{k},\tau)P_l(\hat{k}\cdotp \hat{n}), 
\end{multline}
where $F_{{\rm RD},l}(\vec{k},\tau)$ are the harmonic expansion coefficients.  
The orthonormality of Legendre polynomials allows the evolution equations to 
be written as
\begin{subequations}
\label{per_rd}
\begin{align}
\label{eq:delta_RD}
\dot{\delta}_{\rm RD}&\;=\;-{2 \over 3}(\dot{h}+2\theta_{\rm RD})+{\rho_{\rm DDM} \over \rho_{\rm RD}}(\delta_{\rm DDM}-\delta_{\rm RD}) \\
\dot{\theta}_{\rm RD}&\;=\;k^2({\delta_{\rm RD} \over 4}-\sigma_{\rm RD})-{\rho_{\rm DDM} \over \rho_{\rm RD}}\theta_{\rm RD}
\label{eq:theta_RD}
\end{align}
\begin{multline}
\dot{\sigma}_{\rm RD}={2 \over 15}(2\theta_{\rm RD}+\dot{h}+6\dot{\eta}-{9 \over 2}kF_{\rm RD,3})\\
-{\rho_{\rm DDM} \over \rho_{\rm RD}}\sigma_{\rm RD} \quad
\end{multline}
\begin{multline}
\label{eq:F_RD}
\dot{F}_{\rm RD}={k \over 2l+1}[lF_{{\rm RD},l-1}-(l+1)F_{{\rm RD},l+1}] \\
  -{\rho_{\rm DDM} \over \rho_{\rm RD}}F_{{\rm RD},l} ,\; l\geq 3
\end{multline}
\end{subequations}
at first order, where $\delta_{\rm RD} \equiv F_{\rm RD,1}$, 
$\theta_{\rm RD} \equiv 3/4kF_{\rm RD,1}$, $\sigma_{\rm RD} \equiv F_{\rm RD,2}$, 
and $h$ is the scalar trace of the metric perturbation, all in 
well-established notation.

We have modified the publicly-available {\tt CMBFAST} code 
of \citet{seljak_etal96} to compute the potential power 
spectra.  As we noted in \S~\ref{section:Weak Lensing}, 
we quote the perturbation equations explicitly in synchronous 
gauge simply because {\tt CMBFAST} is written in terms of the 
synchronous gauge perturbations.  Gauge transformations 
can be made straightforwardly \citep{ma_bertschinger95}.

The growth of perturbation is affected by the change of energy density among 
the relativistic and nonrelativistic components. From Eq.~(\ref{rho_ddm}) and 
Eq.~(\ref{rho_rd}) we can see that in the decaying dark matter scenario the 
comoving dark matter density decreases exponentially, and all of this decrement is 
transferred into relativistic energy density.  Consequently, perturbation 
growth exhibits a scale-dependent suppression relative to stable dark matter, 
where the relevant scale is the horizon size at the epoch of decay.  
This late-time suppression of structure growth in large part provides 
the necessary leverage for weak lensing constraints on unstable dark matter.  
For daughter particles, the additional $\rho_{DDM}/ \rho_{RD}$ terms
have an impact on scales greater than the horizon at the 
time of decay \cite{Kaplinghat_etal99}.

As we will discuss below, some of the constraining power 
of weak gravitational lensing comes from observations 
made on scales where linear perturbation theory is 
no longer adequate 
(e.g., \citep{white_hu00,cooray_etal00,white04,huterer_takada05,schmidt08,zentner_etal08}).  
The constraints we forecast in the following sections that are 
based on linear scales only are robust and interesting in 
and of themselves.  However, the utility of weak lensing 
is greatly increased if scales modified by nonlinearity can 
also be exploited for cosmological constraints \citep{zentner_etal08}, 
so we explore multiple proposed nonlinear corrections to linear evolution 
in the following section.

\section{Nonlinear Evolution}
\label{section:Nonlinear}

In the standard application of weak lensing to 
constrain dark energy, most of the constraining power 
comes from scales on which linear evolution of cosmological 
perturbations is no longer valid.  Accounting for nonlinear 
evolution enables a larger range of multipoles to be used, 
and nonlinear evolution greatly enhances signal-to-noise of 
weak lensing measurements at multipoles $\ell \gsim 300$.  
To estimate the constraints that may be anticipated from a full, 
nonlinear treatment of DDM, we explore nonlinear corrections 
to linear evolution using both the method of 
\citet{smith_etal03}, and a halo model-inspired method 
by \citet{peter10}.  \citet{smith_etal03} provide an 
empirical fit for nonlinear power given a linear power 
spectrum.  We utilize the fit of \citet{smith_etal03} directly 
as one of our nonlinear structure models.  This is not entirely 
unreasonable, because lifetimes of interest are far larger than 
a Hubble time, so little decay occurs relative to a standard 
cosmological model.  We implement the 
method of \citet{peter10} using the halo model as follows.

The halo model (see Ref.~\citep{cooray_sheth02} for a review) 
is based on the assumption that all matter resides within 
dark matter halos. The matter power spectrum is given by the sum of two terms,
\beq
\label{eq:P_HM}
P(k)=P_{1H}(k)+P_{2H}(k),
\eeq
where 
\beq
\label{eq:P1}
P_{1H}(k)={1\over\rho_M^2}\int{dm\, m^2 {dn\over dm} \lambda^2(k|m)},
\eeq
and
\beq
\label{eq:P12}
P_{2H}(k)={1\over\rho_M^2} P^{\rm lin}(k)\left[ \int{dm\, m {dn\over dm}\, \lambda(k|m)\, b_h(m) }\right]^2 .
\eeq
In the foregoing equations, $\rho_m$ is the mean matter density of the universe, 
$m$ is halo mass, $\lambda(k|m)$ is the Fourier transform of the Navarro, 
Frenk, and White (NFW, Ref.~\cite{navarro_etal97}) density profile for a halo of 
mass $m$, $P^{\rm lin}(k)$ is the linear matter power spectrum, 
and $b_h(m)$ is the halo bias function.
The one-halo term $P_{1H}(k)$, describes correlations among mass elements within a common 
halo while the two-halo term $P_{2H}(k)$, is due to correlations among mass elements 
in distinct halos.

To estimate the impact of decaying dark matter on matter clustering, we follow 
the approach denoted as \textit{Case 1} by Ref.~\citep{peter10} to describe 
modifications to the halo mass function, halo bias, and internal halo structure.  We 
then incorporate these modifications into the halo model formulae of Eq.~(\ref{eq:P1}) 
and Eq.~(\ref{eq:P12}) to compute lensing power spectra.  This model is based 
upon the assumption that halos at early times are very much like their counterparts 
in models of stable, cold dark matter (because little decay will occur in any 
viable model) and that modifications to halo structure can be described by 
the conservation of adiabatic invariants describing dark matter particle orbits.

Consider a population of dark matter halos that formed prior to any significant dark matter 
decays such that halos at any time $t \ll H_0^{-1}$ can be modeled as standard, 
CDM halos.  These halos then lose mass as their constituent dark matter particles decay. 
If the decay lifetime is much larger than the halo dynamical timescale (as it will always be 
in cases of interest because dynamical times are $\tau_{\rm dyn} \lesssim 0.1 H_0^{-1}$ and viable 
regions of parameter space are $\tau_{\rm DDM} \gg H_0^{-1}$), then 
the halo gravitational potential changes adiabatically.  
Exploiting the adiabatic invariance of angular momentum for particles on 
nearly circular orbits, establishes a prediction for the relationship between 
the initial and final matter distribution within a dark matter halo, 
\beq
\label{eq:ac}
M_{i}(r_{i})r_{i}=M_{f}(r_{f})r_{f},
\eeq
where $M_i(r)$ is the mass enclosed within radius $r$ in the initial, early-time halo, 
$M_f(r)$ is the corresponding quantity describing the contemporary, late-time halo, 
and $r_{i}$ and $r_{f}$ are the initial and final radii of a particle shell, 
assuming that mass shells never cross and particles move in circular orbits.  
Eq.~(\ref{eq:ac}) is the basic relation of the standard, adiabatic contraction 
model for predicting modifications of halo structure due to collisional processes 
\cite{zeldovich_etal80,blumenthal_etal86,gnedin_etal04}.

For unstable dark matter, with a lifetime $\tau_{ddm}$, a fraction $f(\tau_{ddm},z)$ of unstable dark matter 
particles will have decayed by redshift $z$. According to the adiabatic contraction model, 
the mass enclosed in $r_f$ will be
\beq
\label{eq:mft}
M_{f}(r_{f})=(1-f(\tau_{ddm},z))M_{i}(r_{i})
\eeq
Inserting Eq.~(\ref{eq:mft}) into Eq.~(\ref{eq:ac}), the 
relationship between the initial and final radii is 
\beq
\label{eq:mf}
r_{f}=r_{i}/(1-f(\tau_{ddm},z)).
\eeq
If we assume that the initial dark matter halos can be well described by 
NFW profiles, the final mass distribution will be
\begin{align}
\label{eq:rho_new}
\rho_{f}(r_{f})&={1 \over 4\pi r_f^2} {dM_f \over dr_f} \\
&={(1-f)^2 \over 4\pi r_f^2} {dM_i \over dr_i} \\
&={(1-f)^4 \rho_s \over \displaystyle{\left({(1-f)r_f \over r_s}\right)\left[1+{ (1-f)r_f \over r_s}\right]^2}}.
\end{align}

We model the initial mass function $dn_i/dm$ and 
halo bias $b_h(m)$ using the relations of Ref.~\citep{sheth_tormen99}.  
This choice is made for convenience because in models with 
stable dark matter, it satisfies the necessary conditions that the 
halo model integrals contain all mass and that the clustering of 
dark matter is unbiased with respect to itself.  
Some definitions of halo virial radii will be altered by decays.  
In order to ensure that all mass remains accounted for, 
we define halos as the mass within virial radii 
fixed to a definition of 200 times the 
average density of the Universe in the absence of decays.  
Thus, virial radii are fixed to be the same as they would be in standard 
CDM, but halo masses are smaller by a factor of $1-f(\tau_{\rm DDM},z)$.  
This definition preserves the convenient properties of the bias and 
mass relations in Ref.~\citep{sheth_tormen99} and is identical to 
their halo definition in the absence of dark matter decays.

The new halo mass function at mass $M_f$ is
\beq
\label{eq:dndm}
{\dd n_f(M_f[M_i],z) \over \dd M_f}={\dd n_i(M_i,z) \over \dd M_i} \left| {\dd M_i \over \dd M_f} \right|
\eeq
and 
\beq
b_h(M_f)=b_h(M_i),
\eeq
where the initial and final masses are related via Eq.~(\ref{eq:mft}).  
In other words, we assume the abundance and clustering to follow 
the abundance and clustering laws for halos of stable dark matter of 
the corresponding masses.  
Notice that the abundance of halos of a given contemporary 
mass $M_f$ is reduced compared to that in a stable dark matter model 
because the final mass reflects the mass loss due to decays and more 
massive halos are intrinsically rare.  Likewise, 
halos of final mass $M_f$ are more strongly clustered 
than their counterparts in stable dark matter scenarios because 
halo bias is an increasing function of mass 
(see Ref.~\cite{zentner07} for the basic theory of the mass 
function and bias).  The halo density profiles also become shallower 
as $r_s$ increases and $\rho_s$ decreases when the decay-induced modifications 
to halo profiles are accounted for.

\begin{figure}[!t]
\includegraphics[height=8cm]{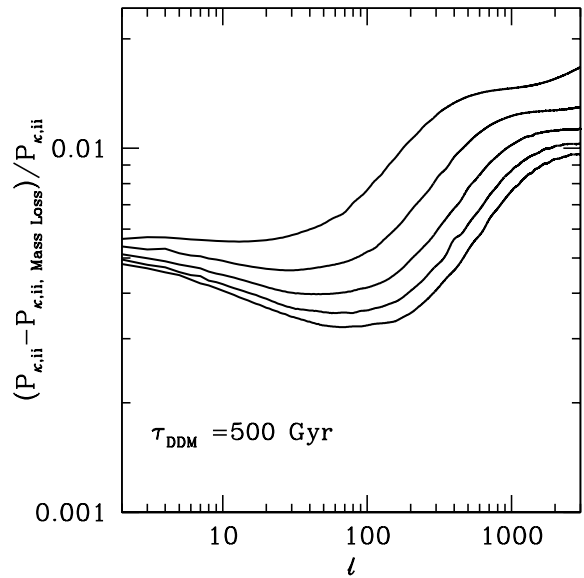}
\caption{
Fractional difference of shear lensing auto-spectra with and without halo 
mass loss due to dark matter decay. In this panel, we choose a dark matter 
lifetime of $\tau=500$~Gyr for illustrative purposes. From top to bottom, 
lines are for the auto-spectra of sources in the first 
tomographic redshift bin ($0 \le z_p < 0.6$) to the 
fifth tomographic redshift bin ($2.4 \le z_p < 3.0$).  In this panel, 
both spectra are computed using a halo model.  For comparison, note that the 
differences between the halo model and the 
Smith et al. \citep{smith_etal03} relation can be 
as much as 20\% with the maximum discrepancy near $\ell \sim 1000$.
}
\label{fig:P_hm}
\end{figure}

The reduction in the number of halos and the shallowing 
of halo profiles reduces lensing power compared 
to a halo model with no accounting for mass loss. Figure~\ref{fig:P_hm} shows a 
comparison between halo model calculations of lensing power spectra 
including and neglecting halo mass loss.  The greatest changes are 
at relatively high $\ell$ ($\ell \gtrsim 300$) and are 
due to the concentration change which alters the one-halo term 
[Eq.~(\ref{eq:P1})].  The shift in the mass function and halo bias 
cause the slight reduction in the two-halo term and power at lower $\ell$. 
As we show in \S~\ref{section:results}, this additional reduction in power 
is a distinctive feature that leads to slightly more restrictive bounds on 
DDM lifetimes.

We emphasize that neither of these approaches 
have been calibrated in detail on simulations of structure 
formation in DDM cosmological models.  
However, we demonstrate that the region of parameter space relevant 
to forthcoming constraints has $\tau_{\rm DDM} \gg H_0^{-1}$ 
(see also \citep{zentner_walker02,Gong_etal08,peter10,peter_etal10}).  
This means that little of the DDM will have 
decayed prior to the present epoch and the boost in signal-to-noise 
should be something close to that afforded by the nonlinear 
treatment of standard, stable dark matter.  In actuality, 
{\em only} a detailed numerical treatment can answer these 
questions definitively (as Refs.~\citep{peter10,peter_etal10}  
have recently argued).  It is our hope that this proof-of-concept 
paper will motivate pursuit of large-scale simulations of DDM 
similar to those being carried out in support of the 
dark energy constraint program 
(e.g., \citep{heitmann_etal05,heitmann_etal08,heitmann_etal09}).

\section{Forecasting Methods}
\label{section:fisher}

The Fisher Information Matrix provides a simple estimate of the parameter 
covariance given data of specified quality.  The Fisher matrix has 
been utilized in numerous, similar contexts in the cosmology literature 
\citep{jungman_etal96,tegmark_etal97,seljak97,hu99,kosowsky_etal02,huterer_takada05,
zentner_etal08,kitching_etal08,peter09,bernstein_huterer10}, 
so we give only a brief review of important results and 
the caveats in our particular application.  

The Fisher matrix of observables in Eq.~(\ref{eq:pkij}), subject to covariance 
as in Eq.~(\ref{cab}), can be written as
\beq
\label{eq:Fij}
F_{{i}{j}}=\sum_{\ell=\ell_{min}}^{\ell_{max}}(2\ell+1)f_{sky}\sum_{A,B} 
\frac{\partial P_{\kappa,A}}{\partial p_{i}}[C^{-1}]_{AB} 
\frac{\partial P_{\kappa,B}}{\partial p_{j}}+F_{ij}^P
\eeq
where the indices A and B run over all $\ntomo (\ntomo + 1)/2$ spectra and 
cross spectra, the $p_{i}$ are the parameters 
of the model, $f_{sky}$ is the fraction of the sky imaged 
by the experiment, and $\ell_{min}=2f_{sky}^{-1/2}$ is the smallest 
multipole constrained by the experiment.  
$F^P_{ij}$ is a prior Fisher matrix 
incorporating previous knowledge of viable regions of parameter space.  
We set $\ell_{max}=3000$ in our most ambitious forecasts.  On 
smaller scales (higher $\ell$), various assumptions such as 
the Gaussianity of the lensing field, break down 
\citep{white_hu00,cooray_hu01,vale_white03,dodelson_etal06,semboloni_etal06,zentner_etal08,rudd_etal08}.  
To be conservative, we explore modest priors to 
each parameter independently, so that 
$F_{ij}^P=\delta_{ij}/(\sigma^P_i)^2$, where 
$\sigma^P_i$ is the 1$\sigma$ prior on parameter $p_{i}$.  
The forecast, 1$\sigma$, marginalized constraint on parameter 
$p_{i}$ is $\sigma(p_{i})=\sqrt{[F^{-1}]_{ii}}$.

Other than the DDM lifetime $\tau_{\rm DDM}$, 
we vary six cosmological parameters that we expect to 
modify weak lensing power spectra at significant levels and 
to exhibit partial degeneracy with $\tau_{\rm DDM}$.  
We construct our forecasts for DDM lifetime bounds after 
marginalizing over the remaining parameters.  Our six additional 
parameters and their fiducial values (in parentheses) are the 
dark energy density $\Omega_{\Lambda}\ (0.74)$, the present-day dark 
matter density, $\omega_{\rm DM}=\Omega_{\rm DM} h^2\ (0.11)$, 
the baryon density $\omega_b=\Omega_b h^2\ (0.023)$, tilt parameter $n_s (0.963)$,
the natural logarithm of the primordial curvature perturbation 
normalization $\ln(\Delta^2_R)\ (-19.94)$, and the sum of the neutrino 
masses $\sum_{i} m_{\nu_i}\ (0.05\ \mathrm{eV})$.  This implies 
a small-scale, low-redshift power spectrum normalization of 
$\sigma_8 \simeq 0.82$.  The optical depth to reionization 
has a negligible effect on the lensing spectra on scales of 
interest, so we do not vary it in our analysis.  
We adopt as our null hypothesis a stable dark matter particle with $\Gamma=0$. 
Note that when $\Gamma \ne 0$, there is a {\em higher} density of 
dark matter in the past than would be inferred for a 
stable dark matter particle because we choose our parameter 
set to describe the contemporary matter density.

We take priors on our cosmological parameters of $\sigma(\omega_m)$ = 0.007, 
$\sigma(\omega_b)$ = $1.2 \times 10^{-3}$, $\sigma(\ln \Delta_R^2)$ = 0.1,
$\sigma(n_s)$ = 0.015, and $\sigma(\Omega_{\Lambda})$ = 0.03. 
We assume no priors on DDM lifetime or 
neutrino mass.  Our fiducial model is motivated by the WMAP seven-year result  
and our priors represent marginalized uncertainties on these parameters based on 
the WMAP seven-year data \citep{Komatsu_etal10}.  These priors are conservative 
and allow for weaker constraints on DDM than would be expected from future data, 
where stronger priors may be available.  To estimate the potential power 
of lensing constraints on DDM when stronger cosmological constraints are 
available, we also explore prior constraints on these parameters 
at the level expected from the Planck mission\footnote{{\tt http://www.esa.int/planck}} 
using the entire Planck prior Fisher matrix of Ref.~\citep{hu_etal06}.  
Of course, using published priors from other analyses is not self-consistent 
because these priors were derived in analyses that assume stable dark 
matter, but relevant lifetimes the dark matter decays should cause only 
subtle alterations to the cosmic microwave background anisotropy spectrum 
so this analysis should approximate a self-consistent simultaneous 
analysis of all data.

In some cases, we will estimate {\em nonlinear} power spectra 
in models with significant neutrino masses.  In such cases, 
we follow the empirical prescription established in previous studies 
(e.g., Refs.~\citep{Hannestad_etal06,kitching_etal08,Ichiki_etal09}) 
and take 
\beq
\label{mnmp}
P_{m}(k)=\left[f_{\nu}\sqrt{P^{\rm lin}_{\nu}(k)}+f_{b+DM}\sqrt{P^{NL}_{b+DM}(k)}\right]^2
\eeq
where 
\begin{subequations}
\label{mne}
\begin{align}
f_{\nu} &= { \Omega_{\nu}\over \Omega_{m}}\mathrm{,}\\
f_{\rm b+DM} &= { \Omega_{\rm DM} + \Omega_{b} \over \Omega_{m}}\mathrm{,}
\end{align}
\end{subequations}
$P^{\rm lin}_{\nu}(k)$ is the linear power spectrum of neutrinos, 
and $P^{\rm NL}_{\rm b+DM}(k)$ is the nonlinear power spectrum 
evaluated for baryons and dark matter only. Again, our adoption 
of this prescription may induce errors in our calculation and 
only a large-scale numerical simulation program can test this 
assumption.

We explore possible constraints from a variety of forthcoming data sets.  
We consider the Dark Energy Survey (DES)\footnote{{\tt http://www.darkenergysurvey.org}} 
as a near-term imaging survey that could provide requisite data for this test.  
We model DES by taking a fractional sky coverage of $f_{sky}=0.12$ and 
with $\bar{n}=15/\mathrm{arcmin}^2$.  Second, we consider a comparably 
narrow, deep imaging survey as may be carried out from a space-based 
platform, such as a Supernova Acceleration Probe-like implementation of a Joint Dark 
Energy Mission (JDEM)\footnote{{\tt http://universe.nasa.gov/program/probes/jdem}}$^{\mathrm{,}}$\footnote{{\tt http://snap.lbl.gov/}} 
or the National Academy of Science's Astronomy and Astrophysics Decadal Survey 
suggestion of a Wide Field Infrared Space Telescope 
(WFIRST)\footnote{{\tt http://sites.nationalacademies.org/bpa/BPA\_049810}}.  
We refer to such a survey as a ``Deep'' survey and 
model it with $f_{sky}=0.05$ and $\bar{n}=100/\mathrm{arcmin}^2$.  
Lastly, we consider a class of future ``Wide'' surveys as may be 
carried out by the Large Synoptic Survey Telescope 
(LSST)\footnote{{\tt http://www.lsst.org}} \citep{lsstbook} or 
Euclid\footnote{{\tt http://sci.esa.int/euclid}} \citep{eicbook}.  
We model these Wide surveys with $f_{sky}=0.5$ and 
$\bar{n}=50/\mathrm{arcmin}^2$.  In all cases, we take 
$\sqrt{\langle \gamma^2 \rangle}=0.2$ and assume particular 
shape measurement errors from each experiment are 
encapsulated in their effective number densities, in accord with 
recent conventional practice in this regard.  Our results are 
relatively insensitive to number density because shot noise 
does not dominate cosmic variance on the scales we consider for 
any of the experimental parameters we consider.

\begin{figure}[!tb]
\includegraphics[height=9cm]{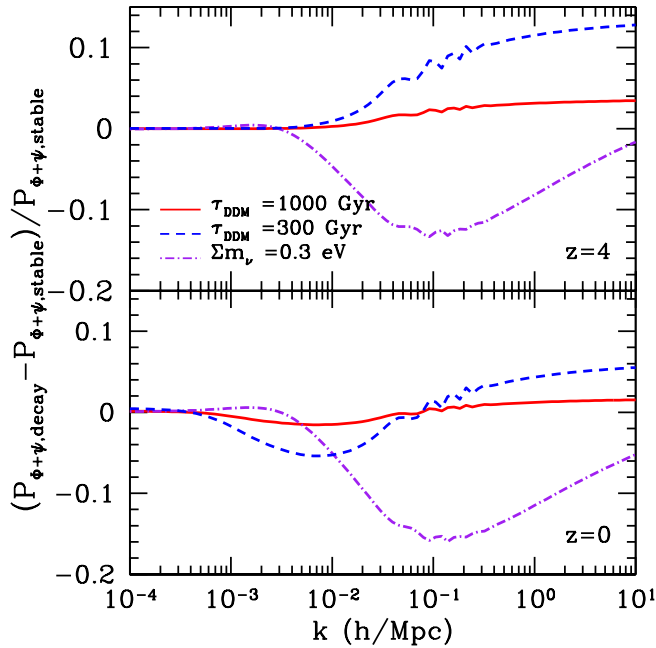}
\caption{
Relative difference in the linear potential power spectra 
between DDM and stable dark matter models at 
$z=4$ ({\em top}) and $z=0$ ({\rm bottom}).  
The {\em solid} lines show a model with 
$\tau_{\rm DDM}=10^3$~Gyr while the {\em dashed} lines 
have $\tau_{\rm DDM}=300$~Gyr.  The {\em dash-dotted} lines 
show the influence of a non-negligible neutrino 
mass with $\sum_{i}\ m_{\nu_i} = 0.3\ \mathrm{eV}$.
}
\label{fig:pps}
\end{figure}

The Fisher matrix is valuable because it greatly reduces the 
computational effort necessary to forecast constraints from 
forthcoming experimental data.  However, the Fisher matrix 
formalism has important drawbacks.  First, the Fisher matrix 
only characterizes parameter degeneracies locally about the 
fiducial model.  Second, the Fisher matrix formalism 
cannot formally be applied to parameters near 
physical limits in their parameter values.  In such cases, 
a Fisher approach allows for parameter degeneracies that 
extend into the forbidden region of the parameter space 
and should not be permitted on physical grounds.  
This additional degeneracy tends to cause under estimates 
of constraints that may be realized from a more detailed analysis.  
An example of this is neutrino mass.  The Fisher matrix has been 
utilized to constrain neutrino mass and empirically 
Fisher matrix projections for neutrino mass constraints 
have been shown to match well direct searches of parameter space 
(e.g., \cite{kitching_etal08,deBernardis_etal09,Ichiki_etal09,debono_etal10,Hannestad_etal06}) 
To verify our Fisher matrix results, we have performed several 
direct searches of subspaces of our full parameter space (limited 
by computational cost) to constrain DDM lifetime.  Generally, 
we find the marginalized Fisher constraints to be only slightly 
less constraining than the direct search results and 
we present an example of this in the following section.  
The computational cost of a full parameter search 
seems unwarranted given the theoretical 
limitations discussed in \S~\ref{section:model}.

\section{Results}
\label{section:results}

\subsection{Weak Lensing Power Spectra}
\label{section:wlpower}

Weak lensing power spectra are altered by DDM in two respects.  
First, the power spectra for potential and density fluctuations 
are altered in a scale-dependent way.  At early epochs, when 
the matter density is higher in the DDM models than in 
standard $\Lambda$CDM, potential and density fluctuations are larger 
because the epoch of matter-radiation equality occurs 
earlier.  We have verified that our constraints are insensitive 
to the epoch at which we normalize the matter density.  
At late times, DDM decays suppress density and 
potential fluctuations.  We show this dependence of potential fluctuations 
on DDM lifetime in Figure~\ref{fig:pps}.  Notice that models of 
unstable dark matter have greater $P_{\Psi-\Phi}(k)$ on scales 
$k \gsim 10^{-2}\ h\ \mathrm{Mpc}^{-1}$ at high redshift, but 
this power is suppressed on subhorizon 
($k \gtrsim 10^{-3}\ h\ \mathrm{Mpc}^{-1}$) 
at lower redshifts.  The strong scale dependence in 
potential power spectra at scales of order 
$k \sim 0.05\ h\ \mathrm{Mpc}^{-1}$ should be 
present in convergence spectra projected on multipoles 
$\ell \sim k\ D_{A}(z=1) \sim 150$ ($z=1$ is the median 
redshift of lensed sources in our model surveys).  The 
different redshift dependence of DDM, which results in 
greater suppression of power with decreasing redshift, 
compared to neutrino mass-induced power suppression allows 
the two to be disentangled.

\begin{figure}[!t]
\includegraphics[height=7.5cm]{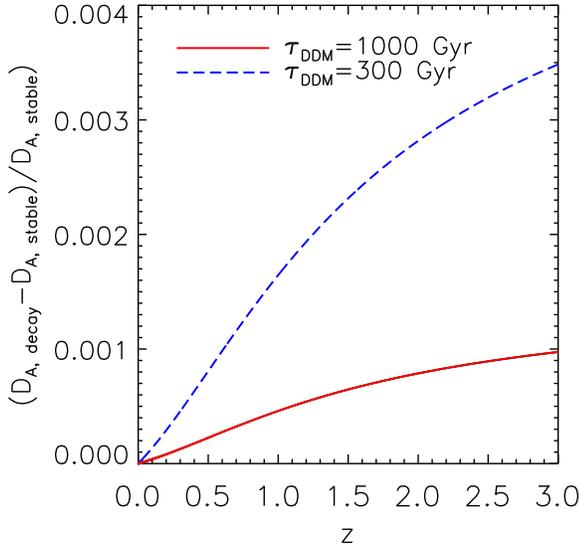}
\caption{
Relative differences in comoving angular diameter 
distance between models with DDM and stable dark 
matter as a function of scale factor.  The {\em solid} 
line represents a model with $\tau_{\rm DDM}=10^3$~Gyr 
and the {\em dashed} line shows a model with 
$\tau_{\rm DDM}=300$~Gyr.}
\label{fig:add}
\end{figure}

The observed strength of gravitational lensing also has a dependence 
upon geometry, so differences in angular diameter distance may lead to 
modified lensing power spectra.  These geometrical differences provide 
the bulk of the information with which lensing can constrain dark 
energy \cite{zhan_knox06,hearin_zentner09} and the angular diameter 
distance to the last-scattering surface has been used to constrain 
decaying {\em dark matter} in previous studies \cite{zentner_walker02,Gong_etal08}.  
In principle, the distance-redshift relation in decaying dark matter 
models can be mimicked by dark energy with a variable equation of 
state \cite{dutta_scherrer10}.  We show in Fig.~\ref{fig:add} 
a comparison of the angular diameter distance in DDM models.  
Fig.~\ref{fig:add} demonstrates the angular diameter distances 
are modified at levels that are small compared to the relative 
potential fluctuations shown in Fig.~\ref{fig:pps}.  As a consequence, 
we find that DDM constraints are based mainly on the scale-dependent 
potential fluctuation modifications, rather than on the modified 
distance scale which is the primary driver of {\em dark energy} 
constraints.

\begin{figure}[!t]
\includegraphics[height=8cm]{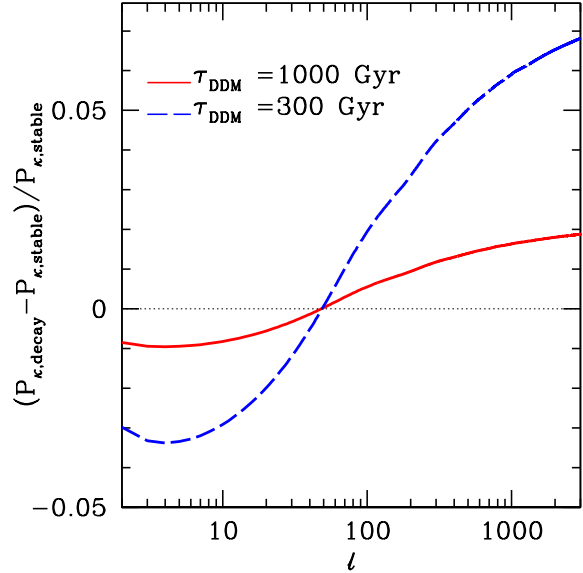}
\caption{
Relative modification to linear lensing convergence power spectra caused by 
DDM.  The lines show the convergence power spectra 
for galaxies in our third redshift bin (galaxies with photometric 
redshifts $1.2 \le z_p < 1.8$) in models with finite $\tau_{\rm DDM}$ 
compared to a model with stable dark matter.  
The {\em solid} line shows the power spectrum 
residual in a model with $\tau_{\rm DDM}=1000$~Gyr, 
and the {\em dashed} line shows the same power spectrum 
residual with $\tau_{\rm DDM}=300$~Gyr.
}
\label{fig:con}
\end{figure}

These changes manifest in a scale- and redshift-dependent shift in 
observable convergence power spectra.  We show examples of the 
shift in the convergence power spectra in models of unstable 
dark matter in Figure~\ref{fig:con}.  Notice the strong scale 
dependence on multipoles of a few hundred.  On smaller 
scales, the net effect is an overall change to the 
level of the convergence power, 
with only a weak scale dependence.  This is one 
reason that we anticipate that we may be able to 
utilize the methods of \citet{smith_etal03} or 
\citet{peter10} to approximate the constraining 
power of weak lensing surveys when nonlinear 
evolution is included.

\subsection{Forecast Constraints on DDM Lifetimes}
\label{section:constraints}

As a first attempt to estimate the power of weak lensing 
to constrain DDM, we examine models in which we consider 
only the linear evolution of potential and density 
fluctuations.  We consider taking the maximum multipole 
that we may observe as high as $\ell_{max}=3000$; however, 
for multipoles larger than $\ell \gsim 300$ nonlinear 
effects will be very important \citep{zentner_etal08} 
(also, see Fig.~\ref{fig:P_hm}).  
Therefore, we also consider taking $\ell_{max}=300$ so 
that we consider only those scales for which linear 
perturbative evolution may be valid.  In this case, 
our theoretical methods are applicable and forthcoming 
weak lensing constraints should do {\em at least} this well.

In our most ambitious forecasts, we 
{\em assume} that we can use the nonlinear 
formula of \citet{smith_etal03} or the halo model 
to estimate the boost in signal that nonlinear evolution 
may provide for weak lensing constraints on DDM.  
These approaches toward nonlinear corrections are not 
entirely self-consistent, but may 
serve as an indicator of what could be achieved 
if a numerical simulation effort addressed nonlinear 
evolution in DDM robustly.

\begin{table}[!t]
\caption{
Forecast 68\% marginalized limits on dark matter decay rates from 
weak lensing surveys under several assumptions.  The 
limits are in units $\Gamma/H_0$, where $H_0=72\ \mathrm{km/s/Mpc}$.   
Constraints are shown for ``Linear'' power spectra, 
``Smith et al.'' nonlinear corrections, 
``Halo Model'' nonlinear corrections, and 
``Modified Halo Model'' nonlinear corrections that 
account for mass loss as in Ref.~\citep{peter10}.  
The abbreviation ``PP'' stands for Planck priors.
}
\vspace*{8pt}
\begin{tabular}{l r r r }
\hline
\hline
Experiment  & DES & Deep & Wide \\ 
\hline
Linear, $\ell_{max}=3000$\ \ \ \ & 0.07\ \ & 0.06\ \ & 0.046\\
Linear, $\ell_{max}=300$\ \ \ \ & 0.08\ \ & 0.09\ \ & 0.057\\
Smith et al, $\ell_{max}=3000$\ \ \ \ & 0.03\ \ & 0.02\ \ & 0.008\\
Smith et al, $\ell_{max}=300$\ \ \ \ & 0.06\ \ & 0.05\ \ & 0.029\\
Halo Model, $\ell_{max}=3000$, \ \ \ \ & 0.03\ \ & 0.02\ \ & 0.010\\
Modified Halo Model, $\ell_{max}=3000$, \ \ \ \ & 0.02\ \ & 0.02\ \ & 0.008\\
\hline
Linear, $\ell_{max}=3000$, PP\ \ \ \ & 0.03\ \ & 0.03\ \ & 0.016\\
Linear, $\ell_{max}=300$, PP\ \ \ \ & 0.06\ \ & 0.07\ \ & 0.026\\
Smith et al, $\ell_{max}=3000$, PP\ \ \ \ & 0.02\ \ & 0.01\ \ & 0.006\\
Smith et al, $\ell_{max}=300$, PP\ \ \ \ & 0.05\ \ & 0.05\ \ & 0.018\\
Halo Model, $\ell_{max}=3000$, PP \ \ \ \ & 0.02\ \ & 0.02\ \ & 0.007\\
Modified Halo Model, $\ell_{max}=3000$, PP\ \ \ \ & 0.02\ \ & 0.01\ \ & 0.006\\
\hline
\end{tabular}
\label{table:limits}
\end{table}

We summarize our primary results for the upper limits 
that may be set on the DDM decay rate $\Gamma$, 
by weak lensing measurements in Table~\ref{table:limits}.  
The limits in this table have been marginalized over 
all other cosmological parameters, including neutrino masses.  
We computed the results in the upper portion of Table~\ref{table:limits} 
using contemporary priors on other cosmological parameters.  
Results below the middle dividing line of Table~\ref{table:limits} 
were computed with prior constraints on cosmology at levels 
expected from the Planck CMB mission and are labeled with 
a ``PP.''  Different lines in Table~\ref{table:limits} show results 
using different model power spectra.  The options are the 
linearly-evolved power spectrum only, results correcting for 
nonlinear evolution using the \citet{smith_etal03} formula, 
nonlinear power results using the halo model, and nonlinear 
power using the halo model modified to account for the loss 
of mass within halos (following Ref.~\citep{peter10}).  
In each case, we consider both restricting to linear scales 
taking $\ell_{\mathrm{max}}=300$ and using nonlinear information 
with $\ell_{\mathrm{max}}=3000$ to constrain decaying dark matter.

Constraints that exploit only linear scales are already promising.  
A DES, a Deep JDEM/WFIRST-like survey, or a Wide LSST- or Euclid-like 
survey should constrain the DDM lifetime at the level of 
$\tau_{\rm DDM} = \Gamma^{-1} \gsim 13 H_0^{-1}$, 
$12 H_0^{-1}$, and $18 H_0^{-1}$ with contemporary priors on 
other cosmological parameters.  These results are already 
comparable to contemporary, model-independent constraints 
on unstable dark matter 
\citep{zentner_walker02,ichiki_etal04,Gong_etal08,Amigo_etal08,peter10,peter_etal10} 
and do not require detailed calibration of nonlinear structure 
growth or ambitious priors on other cosmological parameters 
($\ln \Delta_R^2$ in particular).  It seems reasonable then, 
that weak gravitational lensing will provide, at minimum, 
a complementary, model-independent technique to constrain DDM 
that is competitive with other, existing techniques.

If we interpret the other entries in Table~\ref{table:limits} 
as possible limits that may be achieved if the necessary nonlinear 
evolution in models of DDM can be calibrated, then the results 
become much more interesting.  Using contemporary priors, the limits 
range between $\tau_{\rm DDM} \gsim 33 H_0^{-1}$ and 
$\tau_{\rm DDM} \gsim 43 H_0^{-1}$ for DES, 
$\tau_{\rm DDM} \gsim 48 H_0^{-1}$ and $\tau_{\rm DDM} \gsim 66 H_0^{-1}$ 
for our Deep survey, and  
$\tau_{\rm DDM} \gsim 100 H_0^{-1}$ and $\tau_{\rm DDM} \gsim 125 H_0^{-1}$ 
for our Wide survey.  The variation between the lower 
values and higher values exhibits the range of possible constraints 
estimated using different nonlinear structure formation prescriptions.  
In all cases, the standard halo model gives the poorest constraint 
and the halo model modified to account for mass loss as the dark 
matter decays, as described in \S~\ref{section:Nonlinear}, 
gives the most stringent constraint.  The ability to exploit 
nonlinear power enables weak lensing to constrain unstable 
dark matter more stringently than contemporary methods using 
contemporary priors.

Our most ambitious constraints come from assuming that Planck 
constraints on cosmological parameters other than $\Gamma$ 
are available and that nonlinear structure formation can be 
calibrated sufficiently to make full use of weak lensing data 
on nonlinear scales alongside Planck priors.  These constraints are 
listed in the lower section of Table~\ref{table:limits}.  
In this most ambitious scenario, the lensing constraints 
on unstable dark matter are $\tau_{\rm DDM} \gsim 50 H_0^{-1}$ 
for DES, $\tau_{\rm DDM} \gsim 100 H_0^{-1}$ for a Deep, JDEM/WFIRST-like 
survey, and $\tau_{\rm DDM} \gsim 170 H_0^{-1}$ for a Wide, 
LSST- or Euclid-like survey.  Under these circumstances, 
weak lensing will provide the most stringent, model-independent 
constraints on the decay lifetime of the dark matter particle.

Numerous observational systematics need to be 
controlled for these instruments to achieve their 
statistical limitations and the theory on nonlinear structure 
growth in models with DDM must also be computed more rigorously.  
However, our ambitious forecast limits exceed contemporary bounds 
on dark matter with invisible decay channels significantly, 
so that even if observational or theoretical systematics 
persist, weak lensing may yet provide the strongest limits 
on DDM.  This opportunity (along with the related 
considerations in \citep{ichiki_etal04,peter10,peter_etal10}) 
provides a strong argument for a large-scale computational program 
to study the nonlinear evolution of density fluctuations in models 
with DDM.

\begin{figure}[!t]
\includegraphics[height=10cm]{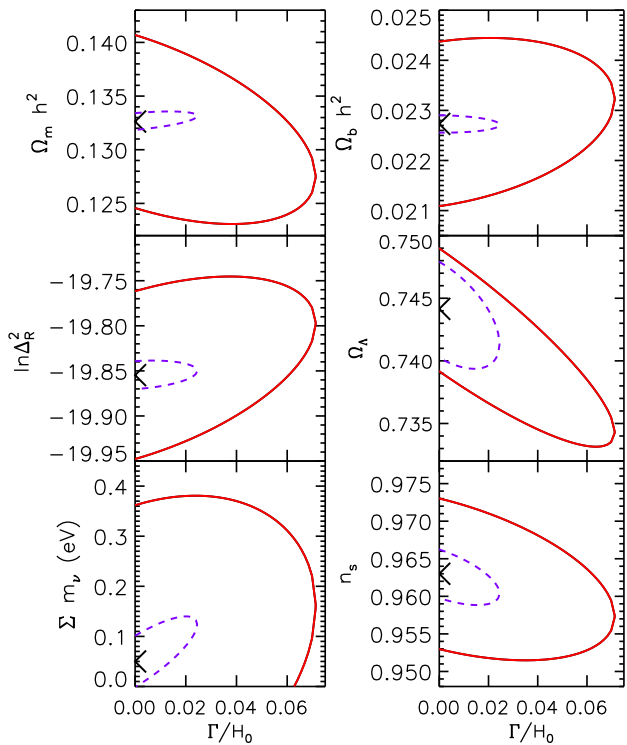}
\caption{
Marginalized, 1$\sigma$ constraint contours for dark matter 
decay rate against other varying cosmological parameters in our 
model.  In this illustrative example, we show contours from our 
linear power spectrum, $\ell_{\mathrm{max}}=300$, Wide survey 
calculations.  The outer contours were computed using our 
contemporary priors while the inner contours were computed 
assuming Planck-level priors.  This example is useful because 
degeneracies are most significant in the linear power spectrum 
calculations.  Notice that the degeneracy between dark matter lifetime 
and neutrino mass is relatively insignificant for contemporary priors, 
but is the most important remaining degeneracy with Planck prior 
constraints.  The crosses designate fiducial parameter values. 
}
\label{fig:contours}
\end{figure}

We conclude our results section, by addressing two remaining outstanding 
issues.  First, the most obvious standard cosmological parameters 
that we may suspect to be degenerate with DDM are neutrino mass and 
the normalization of the matter power spectrum on large scales.  
Much of the constraining power of lensing comes from comparing 
contemporary potential fluctuations to those measured using the 
CMB at high redshift whereas significant neutrino mass also gives 
rise to scale-dependent suppression of the potential power 
spectra.  Figure~\ref{fig:pps} 
provides anecdotal evidence that neutrino mass and 
DDM should not be so degenerate as to destroy constraining power 
because they change the linearly-evolved 
potential spectra in different ways.  Most importantly, 
the effect of DDM is strongly redshift dependent.  
Figure~\ref{fig:contours} shows projected confidence contours 
projected onto two-dimensional subspaces of our parameter space.  
Figure~\ref{fig:contours} shows that significant degeneracies do 
exist when only contemporary priors are used, and Planck priors 
suffice to break most of these degeneracies, leaving 
a slight degeneracy with neutrino mass as the most prominent.  
In the limit of a known neutrino mass (perhaps constrained by 
laboratory experiments), our constraints on DDM lifetimes improve 
by roughly $\sim 25\%$, yielding a best-case constraint from a WIDE survey 
of $\tau_{\rm DDM} \gsim 210 H_0^{-1}$.

Finally, we have performed several direct searches through reduced 
cosmological and DDM parameter space to support our use of the 
Fisher matrix approach in the full parameter space.  The need 
to reduce the parameter space in direct searches is to limit 
computational effort.  At present, with the nonlinear evolution 
in DDM models still uncertain, it does not seem fruitful to 
spend significant computational effort on forecasting.  In a 
simple, linear power model in which we scan the parameter space of 
$\Gamma$, $\omega_m$, $\omega_b$, $ln(\Delta_R^2)$, and $\Omega_{\Lambda}$, 
we find that our direct search decay rate bounds agree with the 
Fisher estimates to within $\sim 20\%$ for our Wide survey. Moreover, the 
Fisher estimates are typically {\em weaker} than the projected direct search limits, 
largely because the Fisher matrix allows for degeneracies among parameters 
that would not be permitted on physical grounds.  Our Fisher matrix 
forecasts are insensitive to the fiducial value of $\Gamma$, so long 
as the DDM lifetime is significantly larger than a Hubble time.

\section{Conclusions}
\label{section:conclusions}

We examined the utility of forthcoming, large-scale imaging surveys to 
constrain the lifetime of dark matter decay into light daughter particles.  
Decaying dark matter can be disentangled from dark energy, because 
its primary observational signature is to reduce the potential fluctuation 
power spectrum, while dark energy is primarily constrained by geometric 
effects \citep{zhan_knox06,hearin_zentner09}.  DDM can be distinguished 
from massive neutrinos because the suppression of power is a strong 
function of redshift.  Assuming a null hypothesis of stable dark matter, we 
found that utilizing only the information from linear scales 
($\ell \lsim 300$) may suffice to place competitive 
limits on dark matter decay rates.  Our linear-only, 
conservative forecast limits are $\Gamma^{-1} \gsim 13H_0^{-1}$ for DES, 
$\Gamma^{-1} \gsim 12H_0^{-1}$ for a Deep JDEM/WFIRST-like survey, and 
$\Gamma^{-1} \gsim 18H_0^{-1}$ for a Wide LSST- or Euclid-like survey.  
The DES limit is slightly weaker than the best contemporary 
constraints \citep{zentner_walker02,Gong_etal08,Amigo_etal08,peter10}, 
while the Wide limit is stronger.  The constraint 
comes largely from a scale- and redshift-dependent reduction in 
convergence power.  This indicates that without any further theoretical 
development, such instruments should provide competitive, 
complementary limits on unstable dark matter.

In practice, lensing surveys will likely measure convergence spectra 
over a wide range of scales where nonlinear evolution is important 
to model ($\ell \sim 10^3$).  The improved signal-to-noise from 
measurement on these scales should dramatically improve upon our 
conservative forecasts.  Assuming that we can use a halo 
model to map our linear spectra onto 
nonlinear spectra, we forecast limits that may be as stringent 
as $\Gamma^{-1} \gsim 50 H_0^{-1}$ for DES, 
$\Gamma^{-1} \gsim 100 H_0^{-1}$ for a Deep JDEM/WFIRST-like survey, 
and $\Gamma^{-1} \gsim 170 H_0^{-1}$ for a Wide LSST- or Euclid-like 
survey when combining with Planck satellite data.  These forecasts are 
more restrictive than constraints available via other means.

Of course, these more aggressive limit forecasts come with caveats.  
Developing surveys must control 
numerous nontrivial systematics to bring their lensing 
programs to fruition and nonlinear evolution in models of 
DDM has not been explored well theoretically.  Weak lensing systematics 
will be explored in support of the well-established goal of using lensing 
to constrain dark energy.  A few groups have performed 
simulations with DDM models to understand its effect on dark matter halos and 
galaxy formation \citep{peter_etal10,Ferrer_etal09}. However the number of 
samples and the simulated halo mass range are not yet sufficient to provide an 
adequate nonlinear fit to matter power spectra.
It is our hope that the opportunity to limit unstable dark matter using information 
on nonlinear scales (see also \citep{peter10}) will motivate 
researchers to explore nonlinear structure formation in models 
of unstable dark matter.  If this can be done, it may be possible 
to limit the dark matter lifetime to be greater than hundreds of 
Hubble times.  Furthermore, it would worthwhile to 
extend such calculations to models with small parent-daughter 
mass splittings.  Such models introduce an additional 
parameter (the mass splitting) that determines the recoil 
energies of the particles after decay.  
Such recoils may alter nonlinear structure at a level 
detectable with lensing data \citep{peter10,peter_etal10}.  
The future is promising for limiting the instability of 
dark matter using forthcoming, large-scale 
astronomical surveys.

\begin{acknowledgments}

We are grateful to Mickey Abbott, Dan Boyanovsky, James Bullock, 
Andrew Hearin, Wayne Hu, Dragan Huterer, Arthur Kosowsky, 
Jeff Newman, Annika Peter, and Chris Purcell for useful comments and 
discussions.  This work was funded by the University of Pittsburgh, the 
National Science Foundation through grant PHY 0968888.  
We thank Uros Seljak and Mathias Zaldarriage 
for use of the publicly-available {\tt CMBFAST} code.  
This work made use of the 
NASA Astrophysics Data System.

\end{acknowledgments}

\bibliography{wldm}

\begin{thebibliography}{89}
\expandafter\ifx\csname natexlab\endcsname\relax\def\natexlab#1{#1}\fi
\expandafter\ifx\csname bibnamefont\endcsname\relax
  \def\bibnamefont#1{#1}\fi
\expandafter\ifx\csname bibfnamefont\endcsname\relax
  \def\bibfnamefont#1{#1}\fi
\expandafter\ifx\csname citenamefont\endcsname\relax
  \def\citenamefont#1{#1}\fi
\expandafter\ifx\csname url\endcsname\relax
  \def\url#1{\texttt{#1}}\fi
\expandafter\ifx\csname urlprefix\endcsname\relax\def\urlprefix{URL }\fi
\providecommand{\bibinfo}[2]{#2}
\providecommand{\eprint}[2][]{\url{#2}}

\bibitem[{\citenamefont{{Jungman}
  et~al.}(1996{\natexlab{a}})\citenamefont{{Jungman}, {Kamionkowski}, and
  {Griest}}}]{jungman_etal96b}
\bibinfo{author}{\bibfnamefont{G.}~\bibnamefont{{Jungman}}},
  \bibinfo{author}{\bibfnamefont{M.}~\bibnamefont{{Kamionkowski}}},
  \bibnamefont{and} \bibinfo{author}{\bibfnamefont{K.}~\bibnamefont{{Griest}}},
  \bibinfo{journal}{\physrep} \textbf{\bibinfo{volume}{267}},
  \bibinfo{pages}{195} (\bibinfo{year}{1996}{\natexlab{a}}),
  \eprint{arXiv:hep-ph/9506380}.

\bibitem[{\citenamefont{{Griest} and
  {Kamionkowski}}(2000)}]{griest_kamionkowski00}
\bibinfo{author}{\bibfnamefont{K.}~\bibnamefont{{Griest}}} \bibnamefont{and}
  \bibinfo{author}{\bibfnamefont{M.}~\bibnamefont{{Kamionkowski}}},
  \bibinfo{journal}{\physrep} \textbf{\bibinfo{volume}{333}},
  \bibinfo{pages}{167} (\bibinfo{year}{2000}).

\bibitem[{\citenamefont{{Bertone} et~al.}(2005)\citenamefont{{Bertone},
  {Hooper}, and {Silk}}}]{bertone_etal05}
\bibinfo{author}{\bibfnamefont{G.}~\bibnamefont{{Bertone}}},
  \bibinfo{author}{\bibfnamefont{D.}~\bibnamefont{{Hooper}}}, \bibnamefont{and}
  \bibinfo{author}{\bibfnamefont{J.}~\bibnamefont{{Silk}}},
  \bibinfo{journal}{\physrep} \textbf{\bibinfo{volume}{405}},
  \bibinfo{pages}{279} (\bibinfo{year}{2005}), \eprint{arXiv:hep-ph/0404175}.

\bibitem[{\citenamefont{{Cooley} and {the CDMS Collaboration}}(2010)}]{cdms10}
\bibinfo{author}{\bibfnamefont{J.}~\bibnamefont{{Cooley}}} \bibnamefont{and}
  \bibinfo{author}{\bibfnamefont{I.}~\bibnamefont{{the CDMS Collaboration}}},
  \bibinfo{journal}{Journal of Physics Conference Series}
  \textbf{\bibinfo{volume}{203}}, \bibinfo{pages}{012004}
  (\bibinfo{year}{2010}), \eprint{arXiv:0912.1601}.

\bibitem[{\citenamefont{{Angle} et~al.}(2008)\citenamefont{{Angle}, {Aprile},
  {Arneodo}, {Baudis}, {Bernstein}, {Bolozdynya}, {Brusov}, {Coelho}, {Dahl},
  {Deviveiros} et~al.}}]{xenon10}
\bibinfo{author}{\bibfnamefont{J.}~\bibnamefont{{Angle}}},
  \bibinfo{author}{\bibfnamefont{E.}~\bibnamefont{{Aprile}}},
  \bibinfo{author}{\bibfnamefont{F.}~\bibnamefont{{Arneodo}}},
  \bibinfo{author}{\bibfnamefont{L.}~\bibnamefont{{Baudis}}},
  \bibinfo{author}{\bibfnamefont{A.}~\bibnamefont{{Bernstein}}},
  \bibinfo{author}{\bibfnamefont{A.}~\bibnamefont{{Bolozdynya}}},
  \bibinfo{author}{\bibfnamefont{P.}~\bibnamefont{{Brusov}}},
  \bibinfo{author}{\bibfnamefont{L.~C.~C.} \bibnamefont{{Coelho}}},
  \bibinfo{author}{\bibfnamefont{C.~E.} \bibnamefont{{Dahl}}},
  \bibinfo{author}{\bibfnamefont{L.}~\bibnamefont{{Deviveiros}}},
  \bibnamefont{et~al.}, \bibinfo{journal}{Physical Review Letters}
  \textbf{\bibinfo{volume}{100}}, \bibinfo{pages}{021303}
  (\bibinfo{year}{2008}), \eprint{arXiv:0706.0039}.

\bibitem[{\citenamefont{{Bernabei} et~al.}(2010)\citenamefont{{Bernabei},
  {Belli}, {Cappella}, {Cerulli}, {Dai}, {d'Angelo}, {He}, {Incicchitti},
  {Kuang}, {Ma} et~al.}}]{dama10}
\bibinfo{author}{\bibfnamefont{R.}~\bibnamefont{{Bernabei}}},
  \bibinfo{author}{\bibfnamefont{P.}~\bibnamefont{{Belli}}},
  \bibinfo{author}{\bibfnamefont{F.}~\bibnamefont{{Cappella}}},
  \bibinfo{author}{\bibfnamefont{R.}~\bibnamefont{{Cerulli}}},
  \bibinfo{author}{\bibfnamefont{C.~J.} \bibnamefont{{Dai}}},
  \bibinfo{author}{\bibfnamefont{A.}~\bibnamefont{{d'Angelo}}},
  \bibinfo{author}{\bibfnamefont{H.~L.} \bibnamefont{{He}}},
  \bibinfo{author}{\bibfnamefont{A.}~\bibnamefont{{Incicchitti}}},
  \bibinfo{author}{\bibfnamefont{H.~H.} \bibnamefont{{Kuang}}},
  \bibinfo{author}{\bibfnamefont{X.~H.} \bibnamefont{{Ma}}},
  \bibnamefont{et~al.}, \bibinfo{journal}{arXiv:1002.1028}
  (\bibinfo{year}{2010}).

\bibitem[{\citenamefont{{Aalseth} et~al.}(2010)\citenamefont{{Aalseth},
  {Barbeau}, {Bowden}, {Cabrera-Palmer}, {Colaresi}, {Collar}, {Dazeley}, {de
  Lurgio}, {Drake}, {Fast} et~al.}}]{cogent10}
\bibinfo{author}{\bibfnamefont{C.~E.} \bibnamefont{{Aalseth}}},
  \bibinfo{author}{\bibfnamefont{P.~S.} \bibnamefont{{Barbeau}}},
  \bibinfo{author}{\bibfnamefont{N.~S.} \bibnamefont{{Bowden}}},
  \bibinfo{author}{\bibfnamefont{B.}~\bibnamefont{{Cabrera-Palmer}}},
  \bibinfo{author}{\bibfnamefont{J.}~\bibnamefont{{Colaresi}}},
  \bibinfo{author}{\bibfnamefont{J.~I.} \bibnamefont{{Collar}}},
  \bibinfo{author}{\bibfnamefont{S.}~\bibnamefont{{Dazeley}}},
  \bibinfo{author}{\bibfnamefont{P.}~\bibnamefont{{de Lurgio}}},
  \bibinfo{author}{\bibfnamefont{G.}~\bibnamefont{{Drake}}},
  \bibinfo{author}{\bibfnamefont{J.~E.} \bibnamefont{{Fast}}},
  \bibnamefont{et~al.}, \bibinfo{journal}{ArXiv e-prints}
  (\bibinfo{year}{2010}), \eprint{arXiv:1002.4703}.

\bibitem[{\citenamefont{{Aprile} and {XENON Collaboration}}(2010)}]{xenon100}
\bibinfo{author}{\bibfnamefont{E.}~\bibnamefont{{Aprile}}} \bibnamefont{and}
  \bibinfo{author}{\bibnamefont{{XENON Collaboration}}},
  \bibinfo{journal}{Journal of Physics Conference Series}
  \textbf{\bibinfo{volume}{203}}, \bibinfo{pages}{012005}
  (\bibinfo{year}{2010}).

\bibitem[{\citenamefont{{Abbasi} et~al.}(2009)}]{icecube09}
\bibinfo{author}{\bibfnamefont{R.}~\bibnamefont{{Abbasi}}}
  \bibnamefont{et~al.}, \bibinfo{journal}{Physical Review Letters}
  \textbf{\bibinfo{volume}{102}}, \bibinfo{pages}{201302}
  (\bibinfo{year}{2009}), \eprint{arXiv:0902.2460}.

\bibitem[{\citenamefont{{Vitale} et~al.}(2009)\citenamefont{{Vitale},
  {Morselli}, and {for the Fermi/LAT Collaboration}}}]{fermi10}
\bibinfo{author}{\bibfnamefont{V.}~\bibnamefont{{Vitale}}},
  \bibinfo{author}{\bibfnamefont{A.}~\bibnamefont{{Morselli}}},
  \bibnamefont{and} \bibinfo{author}{\bibnamefont{{for the Fermi/LAT
  Collaboration}}}, \bibinfo{journal}{arXiv:0912.3828}  (\bibinfo{year}{2009}).

\bibitem[{\citenamefont{{Komatsu} et~al.}(2009)\citenamefont{{Komatsu},
  {Dunkley}, {Nolta}, {Bennett}, {Gold}, {Hinshaw}, {Jarosik}, {Larson},
  {Limon}, {Page} et~al.}}]{Komatsu_etal09}
\bibinfo{author}{\bibfnamefont{E.}~\bibnamefont{{Komatsu}}},
  \bibinfo{author}{\bibfnamefont{J.}~\bibnamefont{{Dunkley}}},
  \bibinfo{author}{\bibfnamefont{M.~R.} \bibnamefont{{Nolta}}},
  \bibinfo{author}{\bibfnamefont{C.~L.} \bibnamefont{{Bennett}}},
  \bibinfo{author}{\bibfnamefont{B.}~\bibnamefont{{Gold}}},
  \bibinfo{author}{\bibfnamefont{G.}~\bibnamefont{{Hinshaw}}},
  \bibinfo{author}{\bibfnamefont{N.}~\bibnamefont{{Jarosik}}},
  \bibinfo{author}{\bibfnamefont{D.}~\bibnamefont{{Larson}}},
  \bibinfo{author}{\bibfnamefont{M.}~\bibnamefont{{Limon}}},
  \bibinfo{author}{\bibfnamefont{L.}~\bibnamefont{{Page}}},
  \bibnamefont{et~al.}, \bibinfo{journal}{\apjs}
  \textbf{\bibinfo{volume}{180}}, \bibinfo{pages}{330} (\bibinfo{year}{2009}),
  \eprint{arXiv:0803.0547}.

\bibitem[{\citenamefont{{Abazajian} et~al.}(2007)\citenamefont{{Abazajian},
  {Markevitch}, {Koushiappas}, and {Hickox}}}]{abazajian_etal07}
\bibinfo{author}{\bibfnamefont{K.~N.} \bibnamefont{{Abazajian}}},
  \bibinfo{author}{\bibfnamefont{M.}~\bibnamefont{{Markevitch}}},
  \bibinfo{author}{\bibfnamefont{S.~M.} \bibnamefont{{Koushiappas}}},
  \bibnamefont{and} \bibinfo{author}{\bibfnamefont{R.~C.}
  \bibnamefont{{Hickox}}}, \bibinfo{journal}{\prd}
  \textbf{\bibinfo{volume}{75}}, \bibinfo{pages}{063511}
  (\bibinfo{year}{2007}), \eprint{arXiv:astro-ph/0611144}.

\bibitem[{\citenamefont{{Randall} et~al.}(2008)\citenamefont{{Randall},
  {Markevitch}, {Clowe}, {Gonzalez}, and {Brada{\v c}}}}]{scott_etal08}
\bibinfo{author}{\bibfnamefont{S.~W.} \bibnamefont{{Randall}}},
  \bibinfo{author}{\bibfnamefont{M.}~\bibnamefont{{Markevitch}}},
  \bibinfo{author}{\bibfnamefont{D.}~\bibnamefont{{Clowe}}},
  \bibinfo{author}{\bibfnamefont{A.~H.} \bibnamefont{{Gonzalez}}},
  \bibnamefont{and} \bibinfo{author}{\bibfnamefont{M.}~\bibnamefont{{Brada{\v
  c}}}}, \bibinfo{journal}{\apj} \textbf{\bibinfo{volume}{679}},
  \bibinfo{pages}{1173} (\bibinfo{year}{2008}), \eprint{arXiv:0704.0261}.

\bibitem[{\citenamefont{{Boyarsky} et~al.}(2009)\citenamefont{{Boyarsky},
  {Lesgourgues}, {Ruchayskiy}, and {Viel}}}]{boyarsky_etal08}
\bibinfo{author}{\bibfnamefont{A.}~\bibnamefont{{Boyarsky}}},
  \bibinfo{author}{\bibfnamefont{J.}~\bibnamefont{{Lesgourgues}}},
  \bibinfo{author}{\bibfnamefont{O.}~\bibnamefont{{Ruchayskiy}}},
  \bibnamefont{and} \bibinfo{author}{\bibfnamefont{M.}~\bibnamefont{{Viel}}},
  \bibinfo{journal}{Journal of Cosmology and Astro-Particle Physics}
  \textbf{\bibinfo{volume}{5}}, \bibinfo{pages}{12} (\bibinfo{year}{2009}),
  \eprint{arXiv:0812.0010}.

\bibitem[{\citenamefont{{Kaplinghat} et~al.}(1999)\citenamefont{{Kaplinghat},
  {Lopez}, {Dodelson}, and {Scherrer}}}]{Kaplinghat_etal99}
\bibinfo{author}{\bibfnamefont{M.}~\bibnamefont{{Kaplinghat}}},
  \bibinfo{author}{\bibfnamefont{R.~E.} \bibnamefont{{Lopez}}},
  \bibinfo{author}{\bibfnamefont{S.}~\bibnamefont{{Dodelson}}},
  \bibnamefont{and} \bibinfo{author}{\bibfnamefont{R.~J.}
  \bibnamefont{{Scherrer}}}, \bibinfo{journal}{\prd}
  \textbf{\bibinfo{volume}{60}}, \bibinfo{pages}{123508}
  (\bibinfo{year}{1999}), \eprint{arXiv:astro-ph/9907388}.

\bibitem[{\citenamefont{{Ichiki} et~al.}(2004)\citenamefont{{Ichiki}, {Oguri},
  and {Takahashi}}}]{ichiki_etal04}
\bibinfo{author}{\bibfnamefont{K.}~\bibnamefont{{Ichiki}}},
  \bibinfo{author}{\bibfnamefont{M.}~\bibnamefont{{Oguri}}}, \bibnamefont{and}
  \bibinfo{author}{\bibfnamefont{K.}~\bibnamefont{{Takahashi}}},
  \bibinfo{journal}{Physical Review Letters} \textbf{\bibinfo{volume}{93}},
  \bibinfo{pages}{071302} (\bibinfo{year}{2004}),
  \eprint{arXiv:astro-ph/0403164}.

\bibitem[{\citenamefont{{Gong} and {Chen}}(2008)}]{Gong_etal08}
\bibinfo{author}{\bibfnamefont{Y.}~\bibnamefont{{Gong}}} \bibnamefont{and}
  \bibinfo{author}{\bibfnamefont{X.}~\bibnamefont{{Chen}}},
  \bibinfo{journal}{\prd} \textbf{\bibinfo{volume}{77}},
  \bibinfo{pages}{023009} (\bibinfo{year}{2008}), \eprint{arXiv:0802.2296}.

\bibitem[{\citenamefont{{Peter}}(2010)}]{peter10}
\bibinfo{author}{\bibfnamefont{A.~H.~G.} \bibnamefont{{Peter}}},
  \bibinfo{journal}{ArXiv e-prints}  (\bibinfo{year}{2010}),
  \eprint{arXiv:1001.3870}.

\bibitem[{\citenamefont{{Peter} et~al.}(2010)\citenamefont{{Peter}, {Moody},
  and {Kamionkowski}}}]{peter_etal10}
\bibinfo{author}{\bibfnamefont{A.~H.~G.} \bibnamefont{{Peter}}},
  \bibinfo{author}{\bibfnamefont{C.~E.} \bibnamefont{{Moody}}},
  \bibnamefont{and}
  \bibinfo{author}{\bibfnamefont{M.}~\bibnamefont{{Kamionkowski}}},
  \bibinfo{journal}{ArXiv e-prints}  (\bibinfo{year}{2010}),
  \eprint{arXiv:1003.0419}.

\bibitem[{\citenamefont{{Zentner} and {Walker}}(2002)}]{zentner_walker02}
\bibinfo{author}{\bibfnamefont{A.~R.} \bibnamefont{{Zentner}}}
  \bibnamefont{and} \bibinfo{author}{\bibfnamefont{T.~P.}
  \bibnamefont{{Walker}}}, \bibinfo{journal}{\prd}
  \textbf{\bibinfo{volume}{65}}, \bibinfo{pages}{063506}
  (\bibinfo{year}{2002}), \eprint{arXiv:astro-ph/0110533}.

\bibitem[{\citenamefont{{De Lope Amigo} et~al.}(2009)\citenamefont{{De Lope
  Amigo}, {Cheung}, {Huang}, and {Ng}}}]{Amigo_etal08}
\bibinfo{author}{\bibfnamefont{S.}~\bibnamefont{{De Lope Amigo}}},
  \bibinfo{author}{\bibfnamefont{W.}~\bibnamefont{{Cheung}}},
  \bibinfo{author}{\bibfnamefont{Z.}~\bibnamefont{{Huang}}}, \bibnamefont{and}
  \bibinfo{author}{\bibfnamefont{S.}~\bibnamefont{{Ng}}},
  \bibinfo{journal}{ArXiv Astrophysics e-prints}  (\bibinfo{year}{2009}),
  \eprint{arXiv:0812.4016}.

\bibitem[{\citenamefont{{Oguri} et~al.}(2003)\citenamefont{{Oguri},
  {Takahashi}, {Ohno}, and {Kotake}}}]{Oguri_etal03}
\bibinfo{author}{\bibfnamefont{M.}~\bibnamefont{{Oguri}}},
  \bibinfo{author}{\bibfnamefont{K.}~\bibnamefont{{Takahashi}}},
  \bibinfo{author}{\bibfnamefont{H.}~\bibnamefont{{Ohno}}}, \bibnamefont{and}
  \bibinfo{author}{\bibfnamefont{K.}~\bibnamefont{{Kotake}}},
  \bibinfo{journal}{\apj} \textbf{\bibinfo{volume}{597}}, \bibinfo{pages}{645}
  (\bibinfo{year}{2003}), \eprint{arXiv:astro-ph/0306020}.

\bibitem[{\citenamefont{{Puetzfeld} and {Chen}}(2004)}]{peutzfeld_chen04}
\bibinfo{author}{\bibfnamefont{D.}~\bibnamefont{{Puetzfeld}}} \bibnamefont{and}
  \bibinfo{author}{\bibfnamefont{X.}~\bibnamefont{{Chen}}},
  \bibinfo{journal}{Classical and Quantum Gravity}
  \textbf{\bibinfo{volume}{21}}, \bibinfo{pages}{2703} (\bibinfo{year}{2004}),
  \eprint{arXiv:gr-qc/0402026}.

\bibitem[{\citenamefont{{Palomares-Ruiz}}(2008)}]{palomares_ruiz08}
\bibinfo{author}{\bibfnamefont{S.}~\bibnamefont{{Palomares-Ruiz}}},
  \bibinfo{journal}{Physics Letters B} \textbf{\bibinfo{volume}{665}},
  \bibinfo{pages}{50} (\bibinfo{year}{2008}), \eprint{arXiv:0712.1937}.

\bibitem[{\citenamefont{{Borzumati} et~al.}(2008)\citenamefont{{Borzumati},
  {Bringmann}, and {Ullio}}}]{borzumati_etal08}
\bibinfo{author}{\bibfnamefont{F.}~\bibnamefont{{Borzumati}}},
  \bibinfo{author}{\bibfnamefont{T.}~\bibnamefont{{Bringmann}}},
  \bibnamefont{and} \bibinfo{author}{\bibfnamefont{P.}~\bibnamefont{{Ullio}}},
  \bibinfo{journal}{\prd} \textbf{\bibinfo{volume}{77}},
  \bibinfo{pages}{063514} (\bibinfo{year}{2008}),
  \eprint{arXiv:hep-ph/0701007}.

\bibitem[{\citenamefont{{Chen} and {Kamionkowski}}(2004)}]{chen_kamionkowski04}
\bibinfo{author}{\bibfnamefont{X.}~\bibnamefont{{Chen}}} \bibnamefont{and}
  \bibinfo{author}{\bibfnamefont{M.}~\bibnamefont{{Kamionkowski}}},
  \bibinfo{journal}{\prd} \textbf{\bibinfo{volume}{70}},
  \bibinfo{pages}{043502} (\bibinfo{year}{2004}),
  \eprint{arXiv:astro-ph/0310473}.

\bibitem[{\citenamefont{{Hansen} and {Haiman}}(2004)}]{hansen_haiman04}
\bibinfo{author}{\bibfnamefont{S.~H.} \bibnamefont{{Hansen}}} \bibnamefont{and}
  \bibinfo{author}{\bibfnamefont{Z.}~\bibnamefont{{Haiman}}},
  \bibinfo{journal}{\apj} \textbf{\bibinfo{volume}{600}}, \bibinfo{pages}{26}
  (\bibinfo{year}{2004}), \eprint{arXiv:astro-ph/0305126}.

\bibitem[{\citenamefont{{Mapelli} et~al.}(2006)\citenamefont{{Mapelli},
  {Ferrara}, and {Pierpaoli}}}]{mapelli_etal06}
\bibinfo{author}{\bibfnamefont{M.}~\bibnamefont{{Mapelli}}},
  \bibinfo{author}{\bibfnamefont{A.}~\bibnamefont{{Ferrara}}},
  \bibnamefont{and}
  \bibinfo{author}{\bibfnamefont{E.}~\bibnamefont{{Pierpaoli}}},
  \bibinfo{journal}{\mnras} \textbf{\bibinfo{volume}{369}},
  \bibinfo{pages}{1719} (\bibinfo{year}{2006}),
  \eprint{arXiv:astro-ph/0603237}.

\bibitem[{\citenamefont{{Zhang} et~al.}(2007)\citenamefont{{Zhang}, {Chen},
  {Kamionkowski}, {Si}, and {Zheng}}}]{zhang_etal07}
\bibinfo{author}{\bibfnamefont{L.}~\bibnamefont{{Zhang}}},
  \bibinfo{author}{\bibfnamefont{X.}~\bibnamefont{{Chen}}},
  \bibinfo{author}{\bibfnamefont{M.}~\bibnamefont{{Kamionkowski}}},
  \bibinfo{author}{\bibfnamefont{Z.}~\bibnamefont{{Si}}}, \bibnamefont{and}
  \bibinfo{author}{\bibfnamefont{Z.}~\bibnamefont{{Zheng}}},
  \bibinfo{journal}{\prd} \textbf{\bibinfo{volume}{76}},
  \bibinfo{pages}{061301} (\bibinfo{year}{2007}), \eprint{0704.2444}.

\bibitem[{\citenamefont{{Y{\"u}ksel} and {Kistler}}(2008)}]{yuksel_kistler08}
\bibinfo{author}{\bibfnamefont{H.}~\bibnamefont{{Y{\"u}ksel}}}
  \bibnamefont{and} \bibinfo{author}{\bibfnamefont{M.~D.}
  \bibnamefont{{Kistler}}}, \bibinfo{journal}{\prd}
  \textbf{\bibinfo{volume}{78}}, \bibinfo{pages}{023502}
  (\bibinfo{year}{2008}), \eprint{arXiv:0711.2906}.

\bibitem[{\citenamefont{{Feldstein} and
  {Fitzpatrick}}(2010)}]{feldstein_fitzpatrick10}
\bibinfo{author}{\bibfnamefont{B.}~\bibnamefont{{Feldstein}}} \bibnamefont{and}
  \bibinfo{author}{\bibfnamefont{A.~L.} \bibnamefont{{Fitzpatrick}}},
  \bibinfo{journal}{arXiv:1003.5662}  (\bibinfo{year}{2010}).

\bibitem[{\citenamefont{{Takahashi} et~al.}(2004)\citenamefont{{Takahashi},
  {Oguri}, and {Ichiki}}}]{Takahashi_etal04}
\bibinfo{author}{\bibfnamefont{K.}~\bibnamefont{{Takahashi}}},
  \bibinfo{author}{\bibfnamefont{M.}~\bibnamefont{{Oguri}}}, \bibnamefont{and}
  \bibinfo{author}{\bibfnamefont{K.}~\bibnamefont{{Ichiki}}},
  \bibinfo{journal}{\mnras} \textbf{\bibinfo{volume}{352}},
  \bibinfo{pages}{311} (\bibinfo{year}{2004}), \eprint{arXiv:astro-ph/0312358}.

\bibitem[{\citenamefont{{Feng} et~al.}(2008)\citenamefont{{Feng}, {Tu}, and
  {Yu}}}]{feng_etal08}
\bibinfo{author}{\bibfnamefont{J.~L.} \bibnamefont{{Feng}}},
  \bibinfo{author}{\bibfnamefont{H.}~\bibnamefont{{Tu}}}, \bibnamefont{and}
  \bibinfo{author}{\bibfnamefont{H.}~\bibnamefont{{Yu}}},
  \bibinfo{journal}{Journal of Cosmology and Astro-Particle Physics}
  \textbf{\bibinfo{volume}{10}}, \bibinfo{pages}{43} (\bibinfo{year}{2008}),
  \eprint{arXiv:0808.2318}.

\bibitem[{\citenamefont{{Feng} et~al.}(2009)\citenamefont{{Feng}, {Kaplinghat},
  {Tu}, and {Yu}}}]{feng_etal09}
\bibinfo{author}{\bibfnamefont{J.~L.} \bibnamefont{{Feng}}},
  \bibinfo{author}{\bibfnamefont{M.}~\bibnamefont{{Kaplinghat}}},
  \bibinfo{author}{\bibfnamefont{H.}~\bibnamefont{{Tu}}}, \bibnamefont{and}
  \bibinfo{author}{\bibfnamefont{H.}~\bibnamefont{{Yu}}},
  \bibinfo{journal}{Journal of Cosmology and Astro-Particle Physics}
  \textbf{\bibinfo{volume}{7}}, \bibinfo{pages}{4} (\bibinfo{year}{2009}),
  \eprint{arXiv:0905.3039}.

\bibitem[{\citenamefont{{Ackerman} et~al.}(2009)\citenamefont{{Ackerman},
  {Buckley}, {Carroll}, and {Kamionkowski}}}]{ackerman_etal09}
\bibinfo{author}{\bibfnamefont{L.}~\bibnamefont{{Ackerman}}},
  \bibinfo{author}{\bibfnamefont{M.~R.} \bibnamefont{{Buckley}}},
  \bibinfo{author}{\bibfnamefont{S.~M.} \bibnamefont{{Carroll}}},
  \bibnamefont{and}
  \bibinfo{author}{\bibfnamefont{M.}~\bibnamefont{{Kamionkowski}}},
  \bibinfo{journal}{\prd} \textbf{\bibinfo{volume}{79}},
  \bibinfo{pages}{023519} (\bibinfo{year}{2009}), \eprint{arXiv:0810.5126}.

\bibitem[{\citenamefont{{Falkowski} et~al.}(2009)\citenamefont{{Falkowski},
  {Juknevich}, and {Shelton}}}]{falkowski_etal09}
\bibinfo{author}{\bibfnamefont{A.}~\bibnamefont{{Falkowski}}},
  \bibinfo{author}{\bibfnamefont{J.}~\bibnamefont{{Juknevich}}},
  \bibnamefont{and}
  \bibinfo{author}{\bibfnamefont{J.}~\bibnamefont{{Shelton}}},
  \bibinfo{journal}{arXiv:0908.1790}  (\bibinfo{year}{2009}).

\bibitem[{\citenamefont{{Kaplan} et~al.}(2009)\citenamefont{{Kaplan}, {Luty},
  and {Zurek}}}]{kaplan_etal09}
\bibinfo{author}{\bibfnamefont{D.~B.} \bibnamefont{{Kaplan}}},
  \bibinfo{author}{\bibfnamefont{M.~A.} \bibnamefont{{Luty}}},
  \bibnamefont{and} \bibinfo{author}{\bibfnamefont{K.~M.}
  \bibnamefont{{Zurek}}}, \bibinfo{journal}{Phys. Rev. D}
  \textbf{\bibinfo{volume}{79}}, \bibinfo{pages}{115016}
  (\bibinfo{year}{2009}).

\bibitem[{\citenamefont{Collaborations}(2009)}]{lsstbook}
\bibinfo{author}{\bibfnamefont{L.~S.} \bibnamefont{Collaborations}},
  \bibinfo{journal}{arXiv:0912.0201}  (\bibinfo{year}{2009}).

\bibitem[{\citenamefont{{Refregier} et~al.}(2010)\citenamefont{{Refregier},
  {Amara}, {Kitching}, {Rassat}, {Scaramella}, {Weller}, and {Euclid Imaging
  Consortium}}}]{eicbook}
\bibinfo{author}{\bibfnamefont{A.}~\bibnamefont{{Refregier}}},
  \bibinfo{author}{\bibfnamefont{A.}~\bibnamefont{{Amara}}},
  \bibinfo{author}{\bibfnamefont{T.~D.} \bibnamefont{{Kitching}}},
  \bibinfo{author}{\bibfnamefont{A.}~\bibnamefont{{Rassat}}},
  \bibinfo{author}{\bibfnamefont{R.}~\bibnamefont{{Scaramella}}},
  \bibinfo{author}{\bibfnamefont{J.}~\bibnamefont{{Weller}}}, \bibnamefont{and}
  \bibinfo{author}{\bibfnamefont{f.~t.} \bibnamefont{{Euclid Imaging
  Consortium}}}, \bibinfo{journal}{ArXiv:1001.0061}  (\bibinfo{year}{2010}).

\bibitem[{\citenamefont{{Schmidt}}(2008)}]{schmidt08}
\bibinfo{author}{\bibfnamefont{F.}~\bibnamefont{{Schmidt}}},
  \bibinfo{journal}{\prd} \textbf{\bibinfo{volume}{78}},
  \bibinfo{pages}{043002} (\bibinfo{year}{2008}), \eprint{0805.4812}.

\bibitem[{\citenamefont{{Ma} et~al.}(2006)\citenamefont{{Ma}, {Hu}, and
  {Huterer}}}]{ma_etal06}
\bibinfo{author}{\bibfnamefont{Z.}~\bibnamefont{{Ma}}},
  \bibinfo{author}{\bibfnamefont{W.}~\bibnamefont{{Hu}}}, \bibnamefont{and}
  \bibinfo{author}{\bibfnamefont{D.}~\bibnamefont{{Huterer}}},
  \bibinfo{journal}{\apj} \textbf{\bibinfo{volume}{636}}, \bibinfo{pages}{21}
  (\bibinfo{year}{2006}), \eprint{astro-ph/0506614}.

\bibitem[{\citenamefont{{Huterer}}(2010)}]{huterer10}
\bibinfo{author}{\bibfnamefont{D.}~\bibnamefont{{Huterer}}},
  \bibinfo{journal}{arXiv:1001.1758}  (\bibinfo{year}{2010}).

\bibitem[{\citenamefont{{Zentner} et~al.}(2008)\citenamefont{{Zentner}, {Rudd},
  and {Hu}}}]{zentner_etal08}
\bibinfo{author}{\bibfnamefont{A.~R.} \bibnamefont{{Zentner}}},
  \bibinfo{author}{\bibfnamefont{D.~H.} \bibnamefont{{Rudd}}},
  \bibnamefont{and} \bibinfo{author}{\bibfnamefont{W.}~\bibnamefont{{Hu}}},
  \bibinfo{journal}{\prd} \textbf{\bibinfo{volume}{77}},
  \bibinfo{pages}{043507} (\bibinfo{year}{2008}), \eprint{arXiv:0709.4029}.

\bibitem[{\citenamefont{{Ma} and {Bertschinger}}(1995)}]{ma_bertschinger95}
\bibinfo{author}{\bibfnamefont{C.}~\bibnamefont{{Ma}}} \bibnamefont{and}
  \bibinfo{author}{\bibfnamefont{E.}~\bibnamefont{{Bertschinger}}},
  \bibinfo{journal}{\apj} \textbf{\bibinfo{volume}{455}}, \bibinfo{pages}{7}
  (\bibinfo{year}{1995}), \eprint{arXiv:astro-ph/9506072}.

\bibitem[{\citenamefont{{Smail}
  et~al.}(1995{\natexlab{a}})\citenamefont{{Smail}, {Hogg}, {Yan}, and
  {Cohen}}}]{smail_etal95a}
\bibinfo{author}{\bibfnamefont{I.}~\bibnamefont{{Smail}}},
  \bibinfo{author}{\bibfnamefont{D.~W.} \bibnamefont{{Hogg}}},
  \bibinfo{author}{\bibfnamefont{L.}~\bibnamefont{{Yan}}}, \bibnamefont{and}
  \bibinfo{author}{\bibfnamefont{J.~G.} \bibnamefont{{Cohen}}},
  \bibinfo{journal}{\apjl} \textbf{\bibinfo{volume}{449}},
  \bibinfo{pages}{L105+} (\bibinfo{year}{1995}{\natexlab{a}}),
  \eprint{arXiv:astro-ph/9506095}.

\bibitem[{\citenamefont{{Smail}
  et~al.}(1995{\natexlab{b}})\citenamefont{{Smail}, {Ellis}, {Fitchett}, and
  {Edge}}}]{smail_etal95b}
\bibinfo{author}{\bibfnamefont{I.}~\bibnamefont{{Smail}}},
  \bibinfo{author}{\bibfnamefont{R.~S.} \bibnamefont{{Ellis}}},
  \bibinfo{author}{\bibfnamefont{M.~J.} \bibnamefont{{Fitchett}}},
  \bibnamefont{and} \bibinfo{author}{\bibfnamefont{A.~C.}
  \bibnamefont{{Edge}}}, \bibinfo{journal}{\mnras}
  \textbf{\bibinfo{volume}{273}}, \bibinfo{pages}{277}
  (\bibinfo{year}{1995}{\natexlab{b}}), \eprint{arXiv:astro-ph/9402049}.

\bibitem[{\citenamefont{{Newman}}(2008)}]{newman08}
\bibinfo{author}{\bibfnamefont{J.~A.} \bibnamefont{{Newman}}},
  \bibinfo{journal}{\apj} \textbf{\bibinfo{volume}{684}}, \bibinfo{pages}{88}
  (\bibinfo{year}{2008}), \eprint{arXiv:0805.1409}.

\bibitem[{\citenamefont{{Bernstein} and {Huterer}}(2010)}]{bernstein_huterer10}
\bibinfo{author}{\bibfnamefont{G.}~\bibnamefont{{Bernstein}}} \bibnamefont{and}
  \bibinfo{author}{\bibfnamefont{D.}~\bibnamefont{{Huterer}}},
  \bibinfo{journal}{\mnras} \textbf{\bibinfo{volume}{401}},
  \bibinfo{pages}{1399} (\bibinfo{year}{2010}), \eprint{arXiv:0902.2782}.

\bibitem[{\citenamefont{{Hearin} et~al.}(2010)\citenamefont{{Hearin},
  {Zentner}, {Ma}, and {Huterer}}}]{hearin_etal10}
\bibinfo{author}{\bibfnamefont{A.~P.} \bibnamefont{{Hearin}}},
  \bibinfo{author}{\bibfnamefont{A.~R.} \bibnamefont{{Zentner}}},
  \bibinfo{author}{\bibfnamefont{Z.}~\bibnamefont{{Ma}}}, \bibnamefont{and}
  \bibinfo{author}{\bibfnamefont{D.}~\bibnamefont{{Huterer}}},
  \bibinfo{journal}{ArXiv e-prints}  (\bibinfo{year}{2010}),
  \eprint{arXiv:1002.3383}.

\bibitem[{\citenamefont{{Massey} et~al.}(2004)\citenamefont{{Massey}, {Rhodes},
  {Refregier}, {Albert}, {Bacon}, {Bernstein}, {Ellis}, {Jain}, {McKay},
  {Perlmutter} et~al.}}]{massey_etal04}
\bibinfo{author}{\bibfnamefont{R.}~\bibnamefont{{Massey}}},
  \bibinfo{author}{\bibfnamefont{J.}~\bibnamefont{{Rhodes}}},
  \bibinfo{author}{\bibfnamefont{A.}~\bibnamefont{{Refregier}}},
  \bibinfo{author}{\bibfnamefont{J.}~\bibnamefont{{Albert}}},
  \bibinfo{author}{\bibfnamefont{D.}~\bibnamefont{{Bacon}}},
  \bibinfo{author}{\bibfnamefont{G.}~\bibnamefont{{Bernstein}}},
  \bibinfo{author}{\bibfnamefont{R.}~\bibnamefont{{Ellis}}},
  \bibinfo{author}{\bibfnamefont{B.}~\bibnamefont{{Jain}}},
  \bibinfo{author}{\bibfnamefont{T.}~\bibnamefont{{McKay}}},
  \bibinfo{author}{\bibfnamefont{S.}~\bibnamefont{{Perlmutter}}},
  \bibnamefont{et~al.}, \bibinfo{journal}{\aj} \textbf{\bibinfo{volume}{127}},
  \bibinfo{pages}{3089} (\bibinfo{year}{2004}),
  \eprint{arXiv:astro-ph/0304418}.

\bibitem[{\citenamefont{{Kasliwal} et~al.}(2008)\citenamefont{{Kasliwal},
  {Massey}, {Ellis}, {Miyazaki}, and {Rhodes}}}]{kasliwal_etal08}
\bibinfo{author}{\bibfnamefont{M.~M.} \bibnamefont{{Kasliwal}}},
  \bibinfo{author}{\bibfnamefont{R.}~\bibnamefont{{Massey}}},
  \bibinfo{author}{\bibfnamefont{R.~S.} \bibnamefont{{Ellis}}},
  \bibinfo{author}{\bibfnamefont{S.}~\bibnamefont{{Miyazaki}}},
  \bibnamefont{and} \bibinfo{author}{\bibfnamefont{J.}~\bibnamefont{{Rhodes}}},
  \bibinfo{journal}{\apj} \textbf{\bibinfo{volume}{684}}, \bibinfo{pages}{34}
  (\bibinfo{year}{2008}), \eprint{0710.3588}.

\bibitem[{\citenamefont{{White} and {Hu}}(2000)}]{white_hu00}
\bibinfo{author}{\bibfnamefont{M.}~\bibnamefont{{White}}} \bibnamefont{and}
  \bibinfo{author}{\bibfnamefont{W.}~\bibnamefont{{Hu}}},
  \bibinfo{journal}{\apj} \textbf{\bibinfo{volume}{537}}, \bibinfo{pages}{1}
  (\bibinfo{year}{2000}).

\bibitem[{\citenamefont{{Cooray} and {Hu}}(2001)}]{cooray_hu01}
\bibinfo{author}{\bibfnamefont{A.}~\bibnamefont{{Cooray}}} \bibnamefont{and}
  \bibinfo{author}{\bibfnamefont{W.}~\bibnamefont{{Hu}}},
  \bibinfo{journal}{\apj} \textbf{\bibinfo{volume}{554}}, \bibinfo{pages}{56}
  (\bibinfo{year}{2001}), \eprint{astro-ph/0012087}.

\bibitem[{\citenamefont{{Vale} and {White}}(2003)}]{vale_white03}
\bibinfo{author}{\bibfnamefont{C.}~\bibnamefont{{Vale}}} \bibnamefont{and}
  \bibinfo{author}{\bibfnamefont{M.}~\bibnamefont{{White}}},
  \bibinfo{journal}{Apj} \textbf{\bibinfo{volume}{592}}, \bibinfo{pages}{699}
  (\bibinfo{year}{2003}).

\bibitem[{\citenamefont{{Dodelson} et~al.}(2006)\citenamefont{{Dodelson},
  {Shapiro}, and {White}}}]{dodelson_etal06}
\bibinfo{author}{\bibfnamefont{S.}~\bibnamefont{{Dodelson}}},
  \bibinfo{author}{\bibfnamefont{C.}~\bibnamefont{{Shapiro}}},
  \bibnamefont{and} \bibinfo{author}{\bibfnamefont{M.}~\bibnamefont{{White}}},
  \bibinfo{journal}{\prd} \textbf{\bibinfo{volume}{73}},
  \bibinfo{pages}{023009} (\bibinfo{year}{2006}),
  \eprint{arXiv:astro-ph/0508296}.

\bibitem[{\citenamefont{{Semboloni} et~al.}(2007)\citenamefont{{Semboloni},
  {van Waerbeke}, {Heymans}, {Hamana}, {Colombi}, {White}, and
  {Mellier}}}]{semboloni_etal06}
\bibinfo{author}{\bibfnamefont{E.}~\bibnamefont{{Semboloni}}},
  \bibinfo{author}{\bibfnamefont{L.}~\bibnamefont{{van Waerbeke}}},
  \bibinfo{author}{\bibfnamefont{C.}~\bibnamefont{{Heymans}}},
  \bibinfo{author}{\bibfnamefont{T.}~\bibnamefont{{Hamana}}},
  \bibinfo{author}{\bibfnamefont{S.}~\bibnamefont{{Colombi}}},
  \bibinfo{author}{\bibfnamefont{M.}~\bibnamefont{{White}}}, \bibnamefont{and}
  \bibinfo{author}{\bibfnamefont{Y.}~\bibnamefont{{Mellier}}},
  \bibinfo{journal}{\mnras} \textbf{\bibinfo{volume}{375}}, \bibinfo{pages}{L6}
  (\bibinfo{year}{2007}), \eprint{arXiv:astro-ph/0606648}.

\bibitem[{\citenamefont{{Seljak} and {Zaldarriaga}}(1996)}]{seljak_etal96}
\bibinfo{author}{\bibfnamefont{U.}~\bibnamefont{{Seljak}}} \bibnamefont{and}
  \bibinfo{author}{\bibfnamefont{M.}~\bibnamefont{{Zaldarriaga}}},
  \bibinfo{journal}{\apj} \textbf{\bibinfo{volume}{469}}, \bibinfo{pages}{437}
  (\bibinfo{year}{1996}), \eprint{arXiv:astro-ph/9603033}.

\bibitem[{\citenamefont{Cooray et~al.}(2000)\citenamefont{Cooray, Hu, and
  Miralda-Escude}}]{cooray_etal00}
\bibinfo{author}{\bibfnamefont{A.}~\bibnamefont{Cooray}},
  \bibinfo{author}{\bibfnamefont{W.}~\bibnamefont{Hu}}, \bibnamefont{and}
  \bibinfo{author}{\bibfnamefont{J.}~\bibnamefont{Miralda-Escude}},
  \bibinfo{journal}{Astrophys. J.} \textbf{\bibinfo{volume}{535}},
  \bibinfo{pages}{L9} (\bibinfo{year}{2000}), \eprint{astro-ph/0003205}.

\bibitem[{\citenamefont{{White}}(2004)}]{white04}
\bibinfo{author}{\bibfnamefont{M.}~\bibnamefont{{White}}},
  \bibinfo{journal}{Astroparticle Physics} \textbf{\bibinfo{volume}{22}},
  \bibinfo{pages}{211} (\bibinfo{year}{2004}), \eprint{astro-ph/0405593}.

\bibitem[{\citenamefont{{Huterer} and {Takada}}(2005)}]{huterer_takada05}
\bibinfo{author}{\bibfnamefont{D.}~\bibnamefont{{Huterer}}} \bibnamefont{and}
  \bibinfo{author}{\bibfnamefont{M.}~\bibnamefont{{Takada}}},
  \bibinfo{journal}{Astroparticle Physics} \textbf{\bibinfo{volume}{23}},
  \bibinfo{pages}{369} (\bibinfo{year}{2005}), \eprint{astro-ph/0412142}.

\bibitem[{\citenamefont{{Smith} et~al.}(2003)\citenamefont{{Smith}, {Peacock},
  {Jenkins}, {White}, {Frenk}, {Pearce}, {Thomas}, {Efstathiou}, and
  {Couchman}}}]{smith_etal03}
\bibinfo{author}{\bibfnamefont{R.~E.} \bibnamefont{{Smith}}},
  \bibinfo{author}{\bibfnamefont{J.~A.} \bibnamefont{{Peacock}}},
  \bibinfo{author}{\bibfnamefont{A.}~\bibnamefont{{Jenkins}}},
  \bibinfo{author}{\bibfnamefont{S.~D.~M.} \bibnamefont{{White}}},
  \bibinfo{author}{\bibfnamefont{C.~S.} \bibnamefont{{Frenk}}},
  \bibinfo{author}{\bibfnamefont{F.~R.} \bibnamefont{{Pearce}}},
  \bibinfo{author}{\bibfnamefont{P.~A.} \bibnamefont{{Thomas}}},
  \bibinfo{author}{\bibfnamefont{G.}~\bibnamefont{{Efstathiou}}},
  \bibnamefont{and} \bibinfo{author}{\bibfnamefont{H.~M.~P.}
  \bibnamefont{{Couchman}}}, \bibinfo{journal}{\mnras}
  \textbf{\bibinfo{volume}{341}}, \bibinfo{pages}{1311} (\bibinfo{year}{2003}),
  \eprint{astro-ph/0207664}.

\bibitem[{\citenamefont{{Cooray} and {Sheth}}(2002)}]{cooray_sheth02}
\bibinfo{author}{\bibfnamefont{A.}~\bibnamefont{{Cooray}}} \bibnamefont{and}
  \bibinfo{author}{\bibfnamefont{R.}~\bibnamefont{{Sheth}}},
  \bibinfo{journal}{\physrep} \textbf{\bibinfo{volume}{372}},
  \bibinfo{pages}{1} (\bibinfo{year}{2002}).

\bibitem[{\citenamefont{{Navarro} et~al.}(1997)\citenamefont{{Navarro},
  {Frenk}, and {White}}}]{navarro_etal97}
\bibinfo{author}{\bibfnamefont{J.~F.} \bibnamefont{{Navarro}}},
  \bibinfo{author}{\bibfnamefont{C.~S.} \bibnamefont{{Frenk}}},
  \bibnamefont{and} \bibinfo{author}{\bibfnamefont{S.~D.~M.}
  \bibnamefont{{White}}}, \bibinfo{journal}{\apj}
  \textbf{\bibinfo{volume}{490}}, \bibinfo{pages}{493} (\bibinfo{year}{1997}),
  \eprint{astro-ph/9611107}.

\bibitem[{\citenamefont{{Zeldovich} et~al.}(1980)\citenamefont{{Zeldovich},
  {Klypin}, {Khlopov}, and {Chechetkin}}}]{zeldovich_etal80}
\bibinfo{author}{\bibfnamefont{Y.}~\bibnamefont{{Zeldovich}}},
  \bibinfo{author}{\bibfnamefont{A.~A.} \bibnamefont{{Klypin}}},
  \bibinfo{author}{\bibfnamefont{M.~Y.} \bibnamefont{{Khlopov}}},
  \bibnamefont{and} \bibinfo{author}{\bibfnamefont{V.~M.}
  \bibnamefont{{Chechetkin}}}, \bibinfo{journal}{Soviet J. Nucl. Phys.}
  \textbf{\bibinfo{volume}{31}}, \bibinfo{pages}{664} (\bibinfo{year}{1980}).

\bibitem[{\citenamefont{{Blumenthal} et~al.}(1986)\citenamefont{{Blumenthal},
  {Faber}, {Flores}, and {Primack}}}]{blumenthal_etal86}
\bibinfo{author}{\bibfnamefont{G.~R.} \bibnamefont{{Blumenthal}}},
  \bibinfo{author}{\bibfnamefont{S.~M.} \bibnamefont{{Faber}}},
  \bibinfo{author}{\bibfnamefont{R.}~\bibnamefont{{Flores}}}, \bibnamefont{and}
  \bibinfo{author}{\bibfnamefont{J.~R.} \bibnamefont{{Primack}}},
  \bibinfo{journal}{\apj} \textbf{\bibinfo{volume}{301}}, \bibinfo{pages}{27}
  (\bibinfo{year}{1986}).

\bibitem[{\citenamefont{{Gnedin} et~al.}(2004)\citenamefont{{Gnedin},
  {Kravtsov}, {Klypin}, and {Nagai}}}]{gnedin_etal04}
\bibinfo{author}{\bibfnamefont{O.~Y.} \bibnamefont{{Gnedin}}},
  \bibinfo{author}{\bibfnamefont{A.~V.} \bibnamefont{{Kravtsov}}},
  \bibinfo{author}{\bibfnamefont{A.~A.} \bibnamefont{{Klypin}}},
  \bibnamefont{and} \bibinfo{author}{\bibfnamefont{D.}~\bibnamefont{{Nagai}}},
  \bibinfo{journal}{\apj} \textbf{\bibinfo{volume}{616}}, \bibinfo{pages}{16}
  (\bibinfo{year}{2004}), \eprint{astro-ph/0406247}.

\bibitem[{\citenamefont{{Sheth} and {Tormen}}(1999)}]{sheth_tormen99}
\bibinfo{author}{\bibfnamefont{R.~K.} \bibnamefont{{Sheth}}} \bibnamefont{and}
  \bibinfo{author}{\bibfnamefont{G.}~\bibnamefont{{Tormen}}},
  \bibinfo{journal}{\mnras} \textbf{\bibinfo{volume}{308}},
  \bibinfo{pages}{119} (\bibinfo{year}{1999}), \eprint{astro-ph/9901122}.

\bibitem[{\citenamefont{{Zentner}}(2007)}]{zentner07}
\bibinfo{author}{\bibfnamefont{A.~R.} \bibnamefont{{Zentner}}},
  \bibinfo{journal}{International Journal of Modern Physics D}
  \textbf{\bibinfo{volume}{16}}, \bibinfo{pages}{763} (\bibinfo{year}{2007}),
  \eprint{arXiv:astro-ph/0611454}.

\bibitem[{\citenamefont{{Heitmann} et~al.}(2005)\citenamefont{{Heitmann},
  {Ricker}, {Warren}, and {Habib}}}]{heitmann_etal05}
\bibinfo{author}{\bibfnamefont{K.}~\bibnamefont{{Heitmann}}},
  \bibinfo{author}{\bibfnamefont{P.~M.} \bibnamefont{{Ricker}}},
  \bibinfo{author}{\bibfnamefont{M.~S.} \bibnamefont{{Warren}}},
  \bibnamefont{and} \bibinfo{author}{\bibfnamefont{S.}~\bibnamefont{{Habib}}},
  \bibinfo{journal}{\apjs} \textbf{\bibinfo{volume}{160}}, \bibinfo{pages}{28}
  (\bibinfo{year}{2005}), \eprint{astro-ph/0411795}.

\bibitem[{\citenamefont{{Heitmann} et~al.}(2008)\citenamefont{{Heitmann},
  {White}, {Wagner}, {Habib}, and {Higdon}}}]{heitmann_etal08}
\bibinfo{author}{\bibfnamefont{K.}~\bibnamefont{{Heitmann}}},
  \bibinfo{author}{\bibfnamefont{M.}~\bibnamefont{{White}}},
  \bibinfo{author}{\bibfnamefont{C.}~\bibnamefont{{Wagner}}},
  \bibinfo{author}{\bibfnamefont{S.}~\bibnamefont{{Habib}}}, \bibnamefont{and}
  \bibinfo{author}{\bibfnamefont{D.}~\bibnamefont{{Higdon}}},
  \bibinfo{journal}{ArXiv:0812.1052}  (\bibinfo{year}{2008}).

\bibitem[{\citenamefont{{Heitmann} et~al.}(2009)\citenamefont{{Heitmann},
  {Higdon}, {White}, {Habib}, {Williams}, and {Wagner}}}]{heitmann_etal09}
\bibinfo{author}{\bibfnamefont{K.}~\bibnamefont{{Heitmann}}},
  \bibinfo{author}{\bibfnamefont{D.}~\bibnamefont{{Higdon}}},
  \bibinfo{author}{\bibfnamefont{M.}~\bibnamefont{{White}}},
  \bibinfo{author}{\bibfnamefont{S.}~\bibnamefont{{Habib}}},
  \bibinfo{author}{\bibfnamefont{B.~J.} \bibnamefont{{Williams}}},
  \bibnamefont{and} \bibinfo{author}{\bibfnamefont{C.}~\bibnamefont{{Wagner}}},
  \bibinfo{journal}{ArXiv:0902.0429}  (\bibinfo{year}{2009}).

\bibitem[{\citenamefont{{Jungman}
  et~al.}(1996{\natexlab{b}})\citenamefont{{Jungman}, {Kamionkowski},
  {Kosowsky}, and {Spergel}}}]{jungman_etal96}
\bibinfo{author}{\bibfnamefont{G.}~\bibnamefont{{Jungman}}},
  \bibinfo{author}{\bibfnamefont{M.}~\bibnamefont{{Kamionkowski}}},
  \bibinfo{author}{\bibfnamefont{A.}~\bibnamefont{{Kosowsky}}},
  \bibnamefont{and} \bibinfo{author}{\bibfnamefont{D.~N.}
  \bibnamefont{{Spergel}}}, \bibinfo{journal}{\prd}
  \textbf{\bibinfo{volume}{54}}, \bibinfo{pages}{1332}
  (\bibinfo{year}{1996}{\natexlab{b}}), \eprint{arXiv:astro-ph/9512139}.

\bibitem[{\citenamefont{{Tegmark} et~al.}(1997)\citenamefont{{Tegmark},
  {Taylor}, and {Heavens}}}]{tegmark_etal97}
\bibinfo{author}{\bibfnamefont{M.}~\bibnamefont{{Tegmark}}},
  \bibinfo{author}{\bibfnamefont{A.~N.} \bibnamefont{{Taylor}}},
  \bibnamefont{and} \bibinfo{author}{\bibfnamefont{A.~F.}
  \bibnamefont{{Heavens}}}, \bibinfo{journal}{\apj}
  \textbf{\bibinfo{volume}{480}}, \bibinfo{pages}{22} (\bibinfo{year}{1997}),
  \eprint{arXiv:astro-ph/9603021}.

\bibitem[{\citenamefont{{Seljak}}(1997)}]{seljak97}
\bibinfo{author}{\bibfnamefont{U.}~\bibnamefont{{Seljak}}},
  \bibinfo{journal}{\apj} \textbf{\bibinfo{volume}{482}}, \bibinfo{pages}{6}
  (\bibinfo{year}{1997}), \eprint{arXiv:astro-ph/9608131}.

\bibitem[{\citenamefont{{Hu}}(1999)}]{hu99}
\bibinfo{author}{\bibfnamefont{W.}~\bibnamefont{{Hu}}},
  \bibinfo{journal}{\apjl} \textbf{\bibinfo{volume}{522}}, \bibinfo{pages}{L21}
  (\bibinfo{year}{1999}), \eprint{astro-ph/9904153}.

\bibitem[{\citenamefont{{Kosowsky} et~al.}(2002)\citenamefont{{Kosowsky},
  {Milosavljevic}, and {Jimenez}}}]{kosowsky_etal02}
\bibinfo{author}{\bibfnamefont{A.}~\bibnamefont{{Kosowsky}}},
  \bibinfo{author}{\bibfnamefont{M.}~\bibnamefont{{Milosavljevic}}},
  \bibnamefont{and}
  \bibinfo{author}{\bibfnamefont{R.}~\bibnamefont{{Jimenez}}},
  \bibinfo{journal}{\prd} \textbf{\bibinfo{volume}{66}},
  \bibinfo{pages}{063007} (\bibinfo{year}{2002}),
  \eprint{arXiv:astro-ph/0206014}.

\bibitem[{\citenamefont{{Kitching} et~al.}(2008)\citenamefont{{Kitching},
  {Heavens}, {Verde}, {Serra}, and {Melchiorri}}}]{kitching_etal08}
\bibinfo{author}{\bibfnamefont{T.~D.} \bibnamefont{{Kitching}}},
  \bibinfo{author}{\bibfnamefont{A.~F.} \bibnamefont{{Heavens}}},
  \bibinfo{author}{\bibfnamefont{L.}~\bibnamefont{{Verde}}},
  \bibinfo{author}{\bibfnamefont{P.}~\bibnamefont{{Serra}}}, \bibnamefont{and}
  \bibinfo{author}{\bibfnamefont{A.}~\bibnamefont{{Melchiorri}}},
  \bibinfo{journal}{\prd} \textbf{\bibinfo{volume}{77}},
  \bibinfo{pages}{103008} (\bibinfo{year}{2008}), \eprint{arXiv:0801.4565}.

\bibitem[{\citenamefont{{Peter}}(2009)}]{peter09}
\bibinfo{author}{\bibfnamefont{A.~H.~G.} \bibnamefont{{Peter}}},
  \bibinfo{journal}{ArXiv e-prints}  (\bibinfo{year}{2009}),
  \eprint{arXiv:0910.4765}.

\bibitem[{\citenamefont{{Rudd} et~al.}(2008)\citenamefont{{Rudd}, {Zentner},
  and {Kravtsov}}}]{rudd_etal08}
\bibinfo{author}{\bibfnamefont{D.~H.} \bibnamefont{{Rudd}}},
  \bibinfo{author}{\bibfnamefont{A.~R.} \bibnamefont{{Zentner}}},
  \bibnamefont{and} \bibinfo{author}{\bibfnamefont{A.~V.}
  \bibnamefont{{Kravtsov}}}, \bibinfo{journal}{\apj}
  \textbf{\bibinfo{volume}{672}}, \bibinfo{pages}{19} (\bibinfo{year}{2008}),
  \eprint{arXiv:astro-ph/0703741}.

\bibitem[{\citenamefont{{Komatsu} et~al.}(2010)\citenamefont{{Komatsu},
  {Smith}, {Dunkley}, {Bennett}, {Gold}, {Hinshaw}, {Jarosik}, {Larson},
  {Nolta}, {Page} et~al.}}]{Komatsu_etal10}
\bibinfo{author}{\bibfnamefont{E.}~\bibnamefont{{Komatsu}}},
  \bibinfo{author}{\bibfnamefont{K.~M.} \bibnamefont{{Smith}}},
  \bibinfo{author}{\bibfnamefont{J.}~\bibnamefont{{Dunkley}}},
  \bibinfo{author}{\bibfnamefont{C.~L.} \bibnamefont{{Bennett}}},
  \bibinfo{author}{\bibfnamefont{B.}~\bibnamefont{{Gold}}},
  \bibinfo{author}{\bibfnamefont{G.}~\bibnamefont{{Hinshaw}}},
  \bibinfo{author}{\bibfnamefont{N.}~\bibnamefont{{Jarosik}}},
  \bibinfo{author}{\bibfnamefont{D.}~\bibnamefont{{Larson}}},
  \bibinfo{author}{\bibfnamefont{M.~R.} \bibnamefont{{Nolta}}},
  \bibinfo{author}{\bibfnamefont{L.}~\bibnamefont{{Page}}},
  \bibnamefont{et~al.}, \bibinfo{journal}{ArXiv e-prints}
  (\bibinfo{year}{2010}), \eprint{arXiv:1001.4538}.

\bibitem[{\citenamefont{{Hu} et~al.}(2006)\citenamefont{{Hu}, {Huterer}, and
  {Smith}}}]{hu_etal06}
\bibinfo{author}{\bibfnamefont{W.}~\bibnamefont{{Hu}}},
  \bibinfo{author}{\bibfnamefont{D.}~\bibnamefont{{Huterer}}},
  \bibnamefont{and} \bibinfo{author}{\bibfnamefont{K.~M.}
  \bibnamefont{{Smith}}}, \bibinfo{journal}{\apjl}
  \textbf{\bibinfo{volume}{650}}, \bibinfo{pages}{L13} (\bibinfo{year}{2006}),
  \eprint{arXiv:astro-ph/0607316}.

\bibitem[{\citenamefont{{Hannestad} et~al.}(2006)\citenamefont{{Hannestad},
  {Tu}, and {Wong}}}]{Hannestad_etal06}
\bibinfo{author}{\bibfnamefont{S.}~\bibnamefont{{Hannestad}}},
  \bibinfo{author}{\bibfnamefont{H.}~\bibnamefont{{Tu}}}, \bibnamefont{and}
  \bibinfo{author}{\bibfnamefont{Y.~Y.} \bibnamefont{{Wong}}},
  \bibinfo{journal}{Journal of Cosmology and Astro-Particle Physics}
  \textbf{\bibinfo{volume}{6}}, \bibinfo{pages}{25} (\bibinfo{year}{2006}),
  \eprint{arXiv:astro-ph/0603019}.

\bibitem[{\citenamefont{{Ichiki} et~al.}(2009)\citenamefont{{Ichiki}, {Takada},
  and {Takahashi}}}]{Ichiki_etal09}
\bibinfo{author}{\bibfnamefont{K.}~\bibnamefont{{Ichiki}}},
  \bibinfo{author}{\bibfnamefont{M.}~\bibnamefont{{Takada}}}, \bibnamefont{and}
  \bibinfo{author}{\bibfnamefont{T.}~\bibnamefont{{Takahashi}}},
  \bibinfo{journal}{\prd} \textbf{\bibinfo{volume}{79}},
  \bibinfo{pages}{023520} (\bibinfo{year}{2009}), \eprint{arXiv:0810.4921}.

\bibitem[{\citenamefont{{de Bernardis} et~al.}(2009)\citenamefont{{de
  Bernardis}, {Kitching}, {Heavens}, and {Melchiorri}}}]{deBernardis_etal09}
\bibinfo{author}{\bibfnamefont{F.}~\bibnamefont{{de Bernardis}}},
  \bibinfo{author}{\bibfnamefont{T.~D.} \bibnamefont{{Kitching}}},
  \bibinfo{author}{\bibfnamefont{A.}~\bibnamefont{{Heavens}}},
  \bibnamefont{and}
  \bibinfo{author}{\bibfnamefont{A.}~\bibnamefont{{Melchiorri}}},
  \bibinfo{journal}{\prd} \textbf{\bibinfo{volume}{80}},
  \bibinfo{pages}{123509} (\bibinfo{year}{2009}), \eprint{arXiv:0907.1917}.

\bibitem[{\citenamefont{{Debono} et~al.}(2010)\citenamefont{{Debono}, {Rassat},
  {R{\'e}fr{\'e}gier}, {Amara}, and {Kitching}}}]{debono_etal10}
\bibinfo{author}{\bibfnamefont{I.}~\bibnamefont{{Debono}}},
  \bibinfo{author}{\bibfnamefont{A.}~\bibnamefont{{Rassat}}},
  \bibinfo{author}{\bibfnamefont{A.}~\bibnamefont{{R{\'e}fr{\'e}gier}}},
  \bibinfo{author}{\bibfnamefont{A.}~\bibnamefont{{Amara}}}, \bibnamefont{and}
  \bibinfo{author}{\bibfnamefont{T.~D.} \bibnamefont{{Kitching}}},
  \bibinfo{journal}{\mnras} pp. \bibinfo{pages}{249--+} (\bibinfo{year}{2010}),
  \eprint{arXiv:0911.3448}.

\bibitem[{\citenamefont{{Zhan} and {Knox}}(2006)}]{zhan_knox06}
\bibinfo{author}{\bibfnamefont{H.}~\bibnamefont{{Zhan}}} \bibnamefont{and}
  \bibinfo{author}{\bibfnamefont{L.}~\bibnamefont{{Knox}}},
  \bibinfo{journal}{ArXiv Astrophysics e-prints}  (\bibinfo{year}{2006}),
  \eprint{arXiv:astro-ph/0611159}.

\bibitem[{\citenamefont{{Hearin} and {Zentner}}(2009)}]{hearin_zentner09}
\bibinfo{author}{\bibfnamefont{A.~P.} \bibnamefont{{Hearin}}} \bibnamefont{and}
  \bibinfo{author}{\bibfnamefont{A.~R.} \bibnamefont{{Zentner}}},
  \bibinfo{journal}{Journal of Cosmology and Astro-Particle Physics}
  \textbf{\bibinfo{volume}{4}}, \bibinfo{pages}{32} (\bibinfo{year}{2009}),
  \eprint{arXiv:0904.3334}.

\bibitem[{\citenamefont{{Dutta} and {Scherrer}}(2010)}]{dutta_scherrer10}
\bibinfo{author}{\bibfnamefont{S.}~\bibnamefont{{Dutta}}} \bibnamefont{and}
  \bibinfo{author}{\bibfnamefont{R.~J.} \bibnamefont{{Scherrer}}},
  \bibinfo{journal}{\prd} \textbf{\bibinfo{volume}{82}},
  \bibinfo{pages}{043526} (\bibinfo{year}{2010}), \eprint{1004.3295}.

\bibitem[{\citenamefont{{Ferrer} et~al.}(2009)\citenamefont{{Ferrer}, {Nipoti},
  and {Ettori}}}]{Ferrer_etal09}
\bibinfo{author}{\bibfnamefont{F.}~\bibnamefont{{Ferrer}}},
  \bibinfo{author}{\bibfnamefont{C.}~\bibnamefont{{Nipoti}}}, \bibnamefont{and}
  \bibinfo{author}{\bibfnamefont{S.}~\bibnamefont{{Ettori}}},
  \bibinfo{journal}{\prd} \textbf{\bibinfo{volume}{80}},
  \bibinfo{pages}{061303} (\bibinfo{year}{2009}), \eprint{arXiv:0905.3161}.

\end{thebibliography}

\end{document}